\documentclass[%
superscriptaddress,
preprint,
5preprintnumbers,
 amsmath,amssymb,
 aps,
]{revtex4-2}
\preprint{APS/123-QED} 
\usepackage[utf8]{inputenc}
\usepackage{hyperref}
\usepackage{amsmath}
\usepackage{amssymb}
\usepackage{bm}
\usepackage{fancyhdr}
\usepackage{pgfplots}
\usepackage{tikz}
\usepackage{caption}
\usepackage{subcaption}
\usepackage{dutchcal}
\usepackage{upgreek}
\usepackage{todonotes}
\usepackage{xcolor}
\usepackage{siunitx}
\usepackage{graphicx}
\usepackage{float}
\usepackage{fancyhdr}

\pgfplotsset{compat=1.16}
\usetikzlibrary{arrows,shapes}
\usetikzlibrary{arrows.meta}

\renewcommand{\paragraph}[1]{{\noindent\textit{#1:}}}

\newcommand{\diff} [1]{\mathrm{d}{#1}}
\newcommand{\grad}{\nabla}    
\newcommand{\intVol}[2]{\int_{#1} #2 \; \diff V}
\newcommand{\intSurf}[2]{\int_{#1} #2 \; \diff S}
\newcommand{\intTime}[3]{\int_{#1}^{#2} #3 \; \diff t}

\newcommand{\auswertungDeltah}{\Tilde{h}}

\newcommand{\trace}[1]{\operatorname{tr}(#1)}

\newcommand{\chr}[1]{ {#1}_\mathrm{chr} }
\newcommand{\dimless}[1]{ {#1}^* }
\newcommand{\charLength}{\chr{l}}
\newcommand{\charVelocity}{\chr{u}}

\newcommand{\Reynolds}{\mathit{Re}}
\newcommand{\Bo}{\mathit{Bo}}  
\newcommand{\Ca}{\mathit{Ca}}  

\newcommand{\gravity}{g}

\definecolor{dgreen}{HTML}{1b9e77}
\definecolor{dred}{HTML}{d95f02}
\definecolor{dblue}{HTML}{377eb8}
\definecolor{dorange}{HTML}{e6ab02}

\pgfplotscreateplotcyclelist{varyPaired}{
  {black},
  {gray},
  {dblue},
  {blue},
  {dred},
  {red},
  {dgreen},
  {green},
  {dorange},
  {orange},
}

\let\originaleps=\epsilon
\let\epsilon=\varepsilon
\let\varepsilon=\originaleps

\DeclareSIUnit\unitVelocity{\metre \per \second}
\DeclareSIUnit\unitVolume{\cubic\metre}
\DeclareSIUnit\unitDensity{\kilogram \per \cubic\metre}
\DeclareSIUnit\unitDynVisc{\kilogram \per \metre \per \second}
\DeclareSIUnit\unitKinVisc{\metre\squared  \per \second}
\DeclareSIUnit\unitStress{\kilogram \per \metre \per \second\squared}
\DeclareSIUnit\unitSurfaceTension{\kilogram \per \second\squared}
\DeclareSIUnit\unitBodyForce{\metre \per \second\squared}
\DeclareSIUnit\unitEnergy{\kilogram\metre\squared \per \second\squared}
\DeclareSIUnit\unitEnergyDensity{\kilogram \per \metre \per \second\squared}
\DeclareSIUnit\unitMobilityCH{\cubic\metre \second \per \kilogram}
\DeclareSIUnit\unitMobilityACNestler{ \metre\squared \second \per \kilogram}
\DeclareSIUnit\unitMobilityAC{ \metre \second \per \kilogram}
\DeclareSIUnit\unitMobilityACtau{\kilogram \per \second \per \metre\squared}
\DeclareSIUnit\unitLaplaceCoeff{\kilogram \metre \per \second\squared}
\DeclareSIUnit\unitPotentialCoeff{\kilogram \per \metre \per \second\squared}

\begin{document}

\title{Modeling the drying process in hard carbon electrodes based on the phase-field method}

\author{M.~Weichel}
\email{marcel.weichel@kit.edu}
\affiliation{
Institute of Nanotechnology (INT), Karlsruhe Institute of Technology, Hermann-von-Helmholtz-Platz 1, 76344 Eggenstein-Leopoldshafen, Germany.
}
\affiliation{%
Institute for Applied Materials (IAM-MMS), Karlsruhe Institute of Technology, Strasse am Forum 7, 76131 Karlsruhe, Germany.
}
\author{M.~Reder}
\affiliation{%
Institute for Applied Materials (IAM-MMS), Karlsruhe Institute of Technology, Strasse am Forum 7, 76131 Karlsruhe, Germany.
}
\affiliation{Institute of Digital Materials Science (IDM), Karlsruhe University of Applied Sciences, Moltkestrasse 30, 76133 Karlsruhe, Germany.}
\author{S.~Daubner}
\affiliation{
Institute of Nanotechnology (INT), Karlsruhe Institute of Technology, Hermann-von-Helmholtz-Platz 1, 76344 Eggenstein-Leopoldshafen, Germany.
}
\affiliation{%
Institute for Applied Materials (IAM-MMS), Karlsruhe Institute of Technology, Strasse am Forum 7, 76131 Karlsruhe, Germany.
}
\author{J.~Klemens}
\affiliation{%
Thin Film Technology (TFT), Karlsruhe Institute of Technology (KIT), Strasse am Forum 7, 76131 Karlsruhe, Germany 
}%
\author{D.~Burger}
\affiliation{%
Thin Film Technology (TFT), Karlsruhe Institute of Technology (KIT), Strasse am Forum 7, 76131 Karlsruhe, Germany 
}%
\author{P.~Scharfer}
\affiliation{%
Thin Film Technology (TFT), Karlsruhe Institute of Technology (KIT), Strasse am Forum 7, 76131 Karlsruhe, Germany 
}%
\author{W.~Schabel}
\affiliation{%
Thin Film Technology (TFT), Karlsruhe Institute of Technology (KIT), Strasse am Forum 7, 76131 Karlsruhe, Germany
}%
\author{B.~Nestler}
\email{britta.nestler@kit.edu}
\affiliation{
Institute of Nanotechnology (INT), Karlsruhe Institute of Technology, Hermann-von-Helmholtz-Platz 1, 76344 Eggenstein-Leopoldshafen, Germany.
}
\affiliation{%
Institute for Applied Materials (IAM-MMS), Karlsruhe Institute of Technology, Strasse am Forum 7, 76131 Karlsruhe, Germany.
}%
\affiliation{Institute of Digital Materials Science (IDM), Karlsruhe University of Applied Sciences, Moltkestrasse 30, 76133 Karlsruhe, Germany.}
\author{D.~Schneider}
\affiliation{
Institute of Nanotechnology (INT), Karlsruhe Institute of Technology, Hermann-von-Helmholtz-Platz 1, 76344 Eggenstein-Leopoldshafen, Germany.
}
\affiliation{%
Institute for Applied Materials (IAM-MMS), Karlsruhe Institute of Technology, Strasse am Forum 7, 76131 Karlsruhe, Germany.
}%
\affiliation{Institute of Digital Materials Science (IDM), Karlsruhe University of Applied Sciences, Moltkestrasse 30, 76133 Karlsruhe, Germany.}

\date{\today}
\begin{abstract}
The present work addresses the simulation of pore emptying during the drying of battery electrodes.
For this purpose, a model based on the multiphase-field method (MPF) is used, since it is an established approach for modeling and simulating multiphysical problems. A model based on phase fields is introduced that takes into account fluid flow, capillary effects, and wetting behavior, all of which play an important role in drying. In addition, the MPF makes it possible to track the movement of the liquid-air interface without computationally expensive adaptive mesh generation. The presented model is used for the first time to investigate pore emptying in real hard carbon microstructures. For this purpose, the microstructures of real dried electrodes are used as input for the simulations. The simulations performed here demonstrate the importance of considering the resolved microstructural information compared to models that rely only on statistical geometry parameters such as pore size distributions. The influence of various parameters such as different microstructures, fluid viscosity, and the contact angle on pore emptying are investigated.
In addition, this work establishes a correlation between the capillary number and the breakthrough time of the solvent as well as the height difference of the solvent front at the time of breakthrough. The results indicate that the drying process can be optimized by doping the particle surface, which changes the contact angle between the fluids and the particles.  
\end{abstract}

\maketitle
\section{Introduction}
The demand for energy storage systems is rising, leading to an increased production of lithium-ion batteries, but post-lithium alternatives such as sodium-ion batteries (SIBs) are also entering the market~\cite{larcher2015towards,usiskin2021fundamentals}. SIBs are a drop-in technology, in the sense that established production processes can be adapted from lithium-ion batteries~\cite{klemens2023challenges}, potentially saving costs and time for upscaling.
However, even with the most modern production processes, there is still potential for a further reduction in energy consumption, which has an impact on costs and sustainability, as drying the battery electrodes in particular is energy intensive~\cite{schutte2024reducing}.
In general, the production of battery electrodes can be divided into five steps: slurry mixing, coating, drying, calendering, and cutting. The slurry consists of active material particles, binding agents, conductive additives, and a solvent. After the slurry is coated onto a substrate, the porous electrode microstructure is formed as a result of the evaporation of the solvent during the drying process. The resulting microstructure affects the electrochemical and mechanical properties of the final battery.
The microstructure formation can be divided into two stages. During the first stage, evaporation of the solvent causes the film height to shrink, resulting in a porous structure, while in the second stage, this porous microstructure is emptied by evaporation.
The distribution of the binder in the microstructure is particularly important for the resulting electrochemical and mechanical properties. It depends on the drying rate, the pore size distribution, and other properties of the slurry. If the drying rate is too high, a migration of binder away from the substrate is observed, which leads to a reduced adhesive strength between the electrode and the substrate. In addition, the accumulation of binder on the electrode surface can lead to pore clogging, which has a negative impact on electrochemical key performance indicators such as rate capability~\cite{westphal2015,jaiser2016investigation,jaiser2017experimental,jaiser2017microstructure,westphal2018critical,kumberg2019drying,kumberg2021influence,kumberg2021investigation,kumberg2021reduced, Klemens2022,klemens2023drying,klemens2023process}.
A detailed understanding of the process-property relationship of drying and the binder distribution offers great potential for battery optimization.

Modeling and simulation can provide important insights into the drying process by predicting the resulting properties from multiple process parameters. The existing modeling approaches for battery production  can be divided into homogenized continuum models~\cite{font2018binder, zihrul2023model}, models based on coarse-grained molecular dynamics~\cite{forouzan2016experiment, liu2023experimentally}, models based on the discrete element method~\cite{lippke2023simulation}, and spatially resolved continuum models~\cite{wolf2024computational,ye2024optimizing}.

Homogenized continuum approaches are based on simplified mathematical equations to investigate the influence of process parameters on drying. Spatial dimensions are often reduced and the flow is calculated according to Darcy's law~\cite{susarla2018modeling}, instead of the Navier-Stokes-equations. Such approaches are less computationally intensive, but at the same time limited due to the underlying assumptions. Investigation of different drying rates have been carried out by Stein et al.~\cite{stein2017mechanistic}. By analyzing the spatial distribution of the additives, they show that a slow two-stage drying process is superior to a fast one-stage drying process. Another mathematical model is introduced by Susarla et al.~\cite{susarla2018modeling}, which is based on volume averaging and the reduction to one dimension, taking into account additional effects of mass and heat transport as well as phase change. By using multi-zone drying, they show that the energy required for drying can be reduced by 50\%. The additives are not taken into account in their simulations.
The work by Font et al.~\cite{font2018binder} shows that the binder distribution is more homogeneous at a low drying rate. The approach of a diffusion-convection equation for the binder is investigated in Zhirul et al.~\cite{zihrul2023model}. Both approaches agree with the trends from experimental observations, but lack the modeling of capillary transport phenomena, which are the main cause of binder migration.
Models based on coarse-grained molecular dynamics are often based on the idea of combining the carbon black and the binder in the CB binder domain (CBD). The evaporation is then modeled by the shrinkage of these CBD particles, since the solvent is contained in them, while  the movement of the solvent is implicitly described by Brownian motion. They are therefore not suitable for simulating the process of pore emptying, which is the subject of this paper. The interactions between the different particles are calculated with coarse-grained molecular dynamics and fitted by a Lennard-Jones potential with shifted force and a granular Hertzian potential. Forouzan et al.~\cite{forouzan2016experiment} describe the simulation of the manufacturing process of lithium-ion cathodes with such an approach on the mesoscale. They investigate numerous physical parameters such as viscosity, shrinkage ratio and the volume fractions of the phases and compare the results with data from experiments to validate their model.
Another work that uses such a modeling approach for the drying step can be found in the ARTISTIC project. It is based on the work of Forouzan et al.~\cite{forouzan2016experiment} and makes it possible to simulate various production steps of the electrode production and electrochemical experiments~\cite{ngandjong2017multiscale,rucci2019tracking,chouchane2019lithium,lombardo2020accelerated,lombardo2021carbon,shodiev20204d,liu2021towards,ngandjong2021investigating,lombardo2022experimentally,liu2023experimentally}. In addition, this model enables the simulation of non-spherical particles~\cite{xu2023lithium}.
Discrete element methods are suitable for describing the particle movement during the first stage of drying and thus simulating the resulting microstructure. In the work by Lippke et al.~\cite{lippke2023simulation}, this method is used to calculate the microstructure formation. In an extension of this work, the discrete element method is extended with further models so that, in addition to film shrinkage, drying kinetics, binder distribution and calendering can also be simulated~\cite{lippke2024drying}. Since discrete element methods use surrogate models for fluid effects, they cannot resolve the drying front and do not take into account the dynamics of the contact lines and thus the effects of surface wettability. They are therefore not suitable for treating the pore emptying process, which is assumed to have the greatest influence on binder migration.

A spatially resolved continuum approach is applied in Wolf et al.~\cite{wolf2024computational}. They model the structural formation during the drying of the battery electrodes. For this purpose, they combine the discrete element method to describe the particle movement and the volume-of-fluid method~\cite{Hirt1981} to distinguish the fluid phases. Scaling is used to keep the simulation time short. However, some physical features that are important for the drying of electrodes are not considered. These are, for example, the wetting behavior including the contact angle, which is important for pore emptying. Additionally, their approach is limited to spherical particles and cannot address the microstructure of non-spherical particles, such as those found in hard carbon electrodes.  
The pore emptying that occurs during the drying of battery electrodes has not been explicitly investigated using the models described above. Models that can take pore emptying into account can be divided into 2 different groups~\cite{lu2021microscale}. On the one hand, there are the discrete pore network models (PNMS) and on the other hand, continuum models (CM)~\cite{whitaker1977simultaneous}. The two model approaches differ in their length scale. PNMS models work on the scale of pores, while continuum models work on the macroscopic length scale. PNMS models are first introduced by Prat et al.~\cite{prat1993percolation} and have been continuously developed since then. For example, properties such as the pore structure and viscosity have been investigated so far~\cite{metzger2005influence, metzger2007influence,metzger2007isothermal, metzger2008viscous}. PNMS are surrogate models and therefore cannot depict spatially resolved microstructures. The information about the microstructures is considered via statistical variables. Furthermore, the computational costs increase if the overall resolution of the simulation scale is increased~\cite{lu2021microscale}. One advantage of homogenized continuum models is that although they can simulate large-scale areas, the influence of the microstructure cannot be taken into account directly, as it only flows into the models via effective parameters. 
When simulating the drying of battery electrodes within the framework of spatially resolved continuum mechanics, several effects must be taken into account, namely the two-phase flow including capillary effects, the movement of the liquid-air interface, and the wetting behavior within complex electrode microstructures. These effects cannot be considered by models based on  homogenized continuum approaches, coarse-grained molecular dynamics or  the discrete element method.
The multiphase-field (MPF) method offers great potential to address these challenges, as it intrinsically covers multiple phases~\cite{steinbach1999generalized, nestler2005multicomponent}, the parametrization of complex microstructures~\cite{reder2023simulative, qu2023multiphysics}, and the evolution of interfaces. Furthermore, coupling with fluid flow has been used in various works~\cite{takaki2018phase, reder2021phase,reder2022phase}, including the simulation of rigid body motion in a fluid~\cite{reder2021phase} and two-phase flows with immersed rigid bodies~\cite{reder2022phase}.
The MPF approach is also well-established in the context of phase transitions and chemical couplings in battery systems~\cite{daubner2022modeling,daubner2024combined}, which simplifies future extensions in terms of model development and multiscale modeling.

In the present work, a model is formulated, validated, and subsequently applied to microstructures of hard carbon electrodes obtained from SEM microscopy data.
The phase-field framework is based on the work of Reder et al.~\cite{reder2022phase}, but includes significant extensions to additionally account for the evaporation of solvent. In Sec.~\ref{sec:mathemtical-formulation}, the model formulation  is presented in detail.
Before we investigate the drying of two hard carbon anode structures for sodium ion batteries in Sec.~\ref{sec:application}, the model is validated based on three example cases (see, Sec.~\ref{sec:validation}).
Together with experimental material parameters, the SEM images are used as an input to study the influence of the drying parameters on the resulting pore emptying behavior. 
This approach provides insights into the spatial velocity field within these complex microstructures, which is related to the binder transport ,and thus, the final binder distribution.
The influence of viscosity, pore size distribution, and the wetting angle between fluid and solid is studied, as these parameters may influence the binder migration. A summary and a discussion of future work can be found in Sec.~\ref{sec:summary_outlook}.
\section{Mathematical Formulation}
\label{sec:mathemtical-formulation}
\begin{figure}[htbp]
\centering
\includegraphics[scale=0.4]{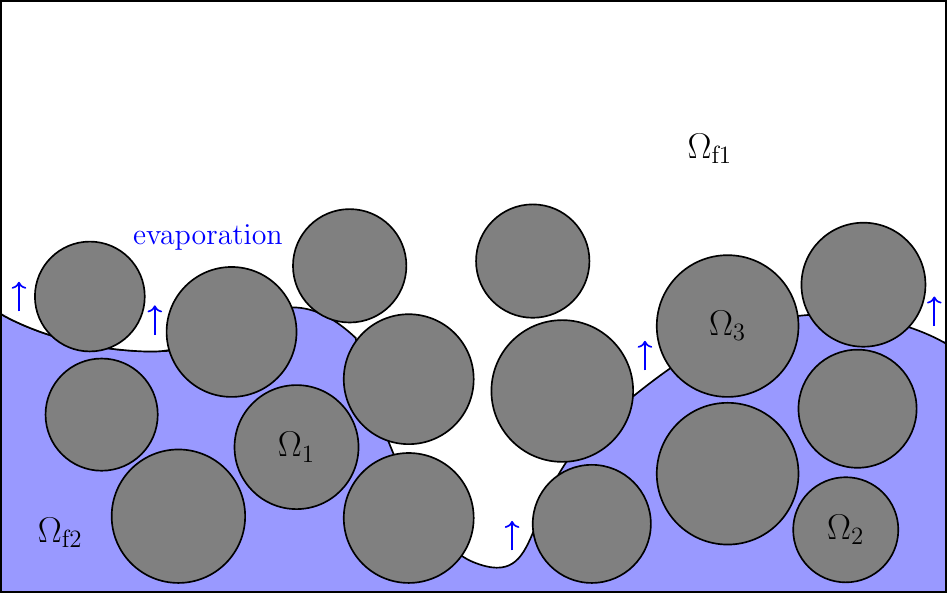}
    \caption{Schematic representation of the drying process of the battery electrodes.}
    \label{fig:schematic-sketch-model}
\end{figure}

\paragraph{Problem Specification}
Fig.~\ref{fig:schematic-sketch-model} shows a schematic representation of the problem addressed in this work.
A multiphase problem is considered, which generally consists of $N^{\text{f}}=2$ different fluid phases and $N^{\text{s}}$ different solid phases. 
The computational domain $\Omega=\Omega_{\text{f}} \cup\Omega_{\text{s}}$ is the union of the domains $ \Omega_{\text{f}} $ and $\Omega_{\text{s}} $, which are occupied by fluid and solid phases, respectively.
Both the fluid and solid domain can be decomposed into the domain of the individual fluid and solid phases, so that $\Omega_{\text{f}} = \Omega_{\text{f1}} \cup \Omega_{\text{f2}}$ and $\Omega_{\text{s}} = \cup_{p=1}^{N^{\text{s}}} \Omega_{p} $
hold.
The model in the present work is based on the approach by Reder et al.~\cite{reder2022phase}, which is adjusted and extended to mimic a phase transformation in the context of battery electrode drying.
First, a model for the two-phase flow assuming $\Omega_{\text{s}} = \emptyset$ is derived, which is based on the approach of Hohenberg et al.~\cite{hohenberg1977theory} (Model-H) and is extended by a term that accounts for the phase transformation in order to mimic evaporation.
This model is then adapted to a diffuse-domain formulation to include solid phases ($\Omega_{\text{s}} \neq \emptyset$).
Therefore, the boundary conditions at the fluid-solid interface are included diffusely, based on the procedure in Li et al.~\cite{li2009solving}.

\paragraph{Two-Phase Flow} Regarding the fluid dynamics, the Navier-Stokes equation is considered, taking the form
\begin{subequations} \label{eq:Navier-Stokes}
  \begin{align}
       \rho \frac{\partial \boldsymbol{u}}{\partial t} + \rho \grad \boldsymbol{u}  \cdot \boldsymbol{u}&= -\grad p + \grad \cdot \left[ \mu \left( \grad\boldsymbol{u} + (\grad \boldsymbol{u})^{\text{T}} \right) \right] + \boldsymbol{K} +\rho \boldsymbol{f}_{\text{V}}, \label{eq:navier-stokes-moment}\\
    \grad \cdot \boldsymbol{u} &= 0 \label{eq:navier-stokes-cont}.
  \end{align}
  \label{eq:nsg}
\end{subequations}
Thereby from equations~\eqref{eq:navier-stokes-moment} and~\eqref{eq:navier-stokes-cont}, the momentum and continuity equations for incompressible flow can be identified. The velocity field is represented by $\bm{u}$, the mass density by $\rho$, the pressure by $p$, the dynamic viscosity by $\mu$, and the body forces by $\bm{f}_{\text{V}}$. The capillary term is denoted as $\boldsymbol{K}$ and will be specified later. 
In general, the dynamic viscosity~$\mu$ can depend on the temperature and, in the context of generalized Newtonian fluids, on the shear rate~cf., e.g.,~\cite{Ruzicka2013}.
The geometry parametrization is achieved using a phase-field method. 
In contrast to a sharp theory, the dividing surfaces between different fluids are replaced by a diffuse transition region, also called diffuse interface (cf.~\cite{reder2024viscous} for a detailed discussion). 
Therefore, phase variables $\varphi_{\alpha}(\bm{x},t) \in [0,1]$ are introduced for each of the $N$ considered phases.
These can be considered as the local volume fraction of a phase $\alpha$ at a specific spatial point~$\bm{x}$ and time~$t$. It follows that the capillary term $\boldsymbol{K} = \boldsymbol{K}(\varphi_1, \varphi_2,\dots,\varphi_N)$ and the physical parameters $\rho = \rho(\varphi_1, \varphi_2,\dots,\varphi_N)$ and $\mu = \mu(\varphi_1, \varphi_2,\dots,\varphi_N)$ depend on the phase variables.\\
For two-phase flow, it is sufficient to consider one phase variable, and so the field
$\Tilde{\varphi} :=\varphi_1 $ is introduced. An Allen-Cahn approach or a Cahn-Hilliard approach can be used to describe the temporal evolution of the phase variables. Details and a comparison can be found in~\cite{Gal2014}. Due to numerical considerations, the Allen-Cahn approach is employed in the present work. It yields
\begin{align}
 \Dot{\Tilde{\varphi}}= \partial_t\Tilde{\varphi} + \grad \Tilde{\varphi} \cdot \bm{u}   =  M  (\beta \partial_{\Tilde{\varphi}}\psi - \alpha  \grad^2 \Tilde{\varphi}) , \qquad  \bm{x} \in \Omega_{{\text{f}}}
\end{align}
with the mobility $M$  and the free energy potential $\psi(\Tilde{\varphi})$. Here, the partial derivative operator with respect to $(\cdot)$ is abbreviated as $\partial_{(\cdot)} := \partial / \partial (\cdot)$. The free energy potential can be selected in various ways. In our case, we use a well potential given by $\psi(\Tilde{\varphi}) = \Tilde{\varphi}^2(1-\Tilde{\varphi})^2$.
The coefficients $\alpha$ and $\beta$ are chosen as $  \alpha = \sigma \epsilon$ and $\beta = 18 \frac{\sigma}{\epsilon}$, where $\sigma$ is the surface energy density and $\epsilon$ scales the diffuse interface width. The value $\epsilon=5\Delta x$ is used in this work.
Moreover, the evolution equation can be written as
\begin{align}
     \Dot{\Tilde{\varphi}} &= M\left[ \frac{36 \sigma}{\epsilon }(2\Tilde{\varphi}^3 -3\Tilde{\varphi}^2 + \Tilde{\varphi})    - \sigma \epsilon \grad^2 \Tilde{\varphi} \right], \qquad  \bm{x} \in \Omega_{\text{f}}.
     \label{eq:allen-cahn-no-evap}
\end{align}
Additionally the wetting boundary condition
\begin{align}
        \alpha \grad \Tilde{\varphi} \cdot \bm{n}^{\text{fs}} &= (\sigma_{\text{2s}}-\sigma_{\text{1s}})\partial_{\Tilde{\varphi}}h^{\text{ff}}(\Tilde{\varphi}), \qquad  \bm{x} \in \partial \Omega_{{\text{f}}}.
        \label{eq:wettingBC}
\end{align}
is considered, where $\bm{n}^{\text{fs}}$ is the outward-pointing normal vector on the boundary of $\Omega_{{\text{f}}}$.
Furthermore, $\sigma_{\text{2s}}$ and $\sigma_{\text{1s}}$ are surface energies between the solid and the respective fluid phase, while $h^{\text{ff}}$ is an interpolation function for the two fluids, which will be specified later.

\paragraph{Capillarity Term}  The capillarity term  $\boldsymbol{K}$ in the Navier-Stokes-equation~\eqref{eq:nsg} accounts for  the curvature minimization due to the surface tension between different fluids. There are various ways to model the capillarity term~\cite{jacqmin1999calculation}. In this work, we take advantage of the so-called potential form
\begin{align}
    \boldsymbol{K}(\Tilde{\varphi}) = -\Tilde{\varphi} \grad \Phi,
\end{align}
where $\Phi =   \beta \partial_{\Tilde{\varphi}}\psi - \alpha  \grad^2 \Tilde{\varphi}$. For a further discussion, see Jacqmin et al.~\cite{jacqmin1996energy}.
The Allen-Cahn approach from equation~\eqref{eq:allen-cahn-no-evap} intrinsically exhibits curvature minimizing dynamics. This leads to undesirable dynamics, as surface tension effects are already covered by the capillary term, which provides a corresponding velocity contribution.
The additional surface-minimizing dynamics in the Allen-Cahn equation can be removed~\cite{Schoof2018}, which is achieved by the approach of Sun et al.~\cite{Sun2007}.
It follows for the phase field equation
\begin{align}
     \Dot{\Tilde{\varphi}} &= M\left[ \frac{36 \sigma}{\epsilon }(2\Tilde{\varphi}^3 -3\Tilde{\varphi}^2 + \Tilde {\varphi}) - \sigma \epsilon( \grad^2 \Tilde{\varphi} - ||\grad\Tilde{\varphi}||\grad \cdot \bm{n} ) \right], \qquad \bm{x} \in \Omega_{\text{f}}
     \label{eq:allen-cahn-surface-min}
\end{align}
with $\bm{n} = \frac{\grad\Tilde{\varphi}}{||\grad\Tilde{\varphi}||}$. This extension ensures that the influence of surface tension is not overestimated. 

\paragraph{Evaporation Term} The present model is intended to mimic the drying process in battery electrode systems.
The evaporation phenomenon can be modeled by a phase transformation.
For this purpose, we insert a term $ v^{\text{e}} ||\grad \Tilde{\varphi}||$ into equation~\eqref{eq:allen-cahn-surface-min} to model the motion of the interface with a phase transformation velocity $v^{\text{e}}$ and a term $ ||\grad \Tilde{\varphi}||$ that distributes the velocity over the diffuse interface.
Thus, equation~\eqref{eq:allen-cahn-surface-min} is modified as follows:
\begin{align}
   \Dot{\Tilde{\varphi}} &= M\left[ \frac{36 \sigma}{\epsilon }(2\Tilde{\varphi}^3 -3\Tilde{\varphi}^2 + \Tilde{\varphi})    - \sigma \epsilon\left(  \grad^2 \Tilde{\varphi} - ||\grad\Tilde{\varphi}||\grad \cdot \bm{n} \right) \right] + v^{\text{e}} ||\grad \Tilde{\varphi}||, \qquad  \bm{x} \in \Omega_{\text{f}}. 
    \label{eq:Allen-Cahn-Gleichung-Evap}
\end{align}
According to the equation~\eqref{eq:kappa-and-evaprate}, the velocity $v^{\text{e}}$ is directly related to the evaporation rate (mass flux per area in \si{\kg \per\metre\squared\per\second }), by means of the factor~$\kappa$ (unit \si{\per\second}). It should be noted that a constant evaporation rate occurs for most of the pore emptying~\cite{jaiser2016investigation}, which means that $\kappa$ is constant over time. By incorporating $\kappa$, we can calculate the theoretical total volume fraction of the solvent
\begin{align}
  \chi_{\text{solvent,theo}}^{t_n} = -\kappa \cdot t_n + \chi_{\text{solvent}}^0,
    \label{eq:evapfromtime}
\end{align}
where $t_n$ is the time and $\chi_{\text{solvent}}^0$ denotes the initial volume fraction of the fluid that evaporates. The derivation of the equation~\eqref{eq:evapfromtime} can be found in the Appendix~\ref{derivation-of-evap}. The velocity $v^{\text{e}}$ then results in
\begin{align}
    v^{\text{e}} = - \frac{\chi_{\text{solvent,theo}}^{t_n} - \chi_{\text{film}}^{t_n}}{\int || \grad \Tilde{\varphi}|| \text{d}V  \cdot \Delta t} V^0_{\text{film}}, 
    \label{eq:volume-velocity}
\end{align}
where $V^0_{\text{film}} $ is the total film volume at $t=0$. The term of the driving force can be interpreted as a Lagrange multiplier, which ensures the target volume fraction $\chi_{\text{solvent,theo}}^{t_n}$.
For the special case $\kappa = 0$, the Allen-Cahn equation is volume-preserving (see Appendix~\ref{volume-perserving-allen-cahn}). 

\paragraph{Multiphase Flow with Rigid Body Coupling}
To apply the Model-H approach~\cite{hohenberg1977theory} to the two-phase flow with solid phases, which exhibit a diffuse solid interface, a normalization approach according to Reder et al.~\cite{reder2022phase} is employed. This normalization splits the problem into two sub-problems. First, we introduce the local volume fraction of all fluid  and solid phases in the form of 
\begin{align}
        \varphi^{\text{f}} = \frac{V^{\text{f}}}{V} = \sum_{\alpha=1}^{N^{\text{f}}}   \varphi_{\alpha}^{\text{f}} \qquad \text{and}  \qquad
            \varphi^{\text{s}} = \frac{V^{\text{s}}}{V} = \sum_{\alpha=1}^{N^{\text{s}}}   \varphi_{\alpha}^{\text{s}},    
\end{align}
with $V^{\text{f}}$ as the total fluid  and $V^{\text{s}}$ as the total solid volume in a representative volume element~$V$. Moreover, the normalization for the fluid phases is done by 
\begin{align}
    \Tilde{\varphi}^{\text{f}}_{\alpha} := \frac{\varphi^{\text{f}}_{\alpha}}{\varphi^{\text{f}}}, \quad \text{for} \quad \varphi^{\text{f}}>0,
\end{align}
where $\Tilde{\varphi} :=\Tilde{\varphi}_1 $, in the case of two fluids, including the sum condition $\sum_{\alpha=1}^{N^{\text{f}}}  \Tilde{\varphi}^{\text{f}}_{\alpha} = 1 $. Regarding the normalized phase variable $\Tilde{\varphi}$, the evolution equation~\eqref{eq:Allen-Cahn-Gleichung-Evap} can be employed.
For the entire set of partial differential equations (PDEs), the appropriate boundary conditions at the fluid-solid interface must be defined via the diffuse interface.
In general, the approach of Li et al.~\cite{li2009solving} can be used to formulate boundary conditions for a diffuse interface between different phases.
The boundary conditions can be of the Robin, Dirichlet, or Neumann type.
With regard to the Navier-Stokes equation system \eqref{eq:Navier-Stokes}, we follow the approach of Beckermann et al.~\cite{beckermann1999modeling} for the diffusive application of the no-slip boundary condition.
As for the wetting boundary condition~\eqref{eq:wettingBC}, the general approach is used, e.g., in ~\cite{Aland2010,reder2022phase} to derive diffuse wetting boundary conditions in a Cahn-Hilliard formulation. The derivation of the corresponding Allen-Cahn formulation can be found in Appendix~\ref{derivation-diffuse-wetting-bc}. Thus, the set of PDEs for the whole domain $ \bm{x} \in \Omega$ that need to be solved is as follows:
\begin{subequations}
\label{eq:allen-cahn-final}
\begin{align}
  \begin{split}
    \partial_t(\rho h^{\text{fs}} \boldsymbol{u}) &= -\grad \cdot (\rho  h^{\text{fs}}  \boldsymbol{u}  \otimes\boldsymbol{u}) -h^{\text{fs}}\grad p \\
    &\qquad + \grad \cdot \left( h^{\text{fs}}\mu  \grad\boldsymbol{u}\right) + \boldsymbol{u} \grad \cdot (\mu \grad h^{\text{fs}}) +\rho h^{\text{fs}}  \boldsymbol{f}_{\text{V}} +  h^{\text{fs}}\boldsymbol{K}
  \end{split} \label{eq:1a} \\
  0 &= \grad \cdot (h^{\text{fs}}\boldsymbol{u}), \label{eq:1b} \\
  \begin{split}
    h^{\text{fs}}  \Dot{\Tilde{\varphi}} &=  M\Big[\beta h^{\text{fs}} \partial_{\Tilde{\varphi}} \psi - \alpha \left(\grad \cdot h^{\text{fs}} \grad \Tilde{\varphi} - h^{\text{fs}} ||\grad\Tilde{\varphi}||\grad \cdot \frac{\grad\Tilde{\varphi}}{||\grad\Tilde{\varphi}||}\right) \\
    &\qquad - \partial_{\varphi^{\text{f}}}h^{\text{fs}}||\grad\varphi^{\text{f}}|| (\sigma_{\text{2s}}-\sigma_{\text{1s}})\partial_{\Tilde{\varphi}}h^{\text{ff}}\Big] + h^{\text{fs}} v^{\text{e}} ||\grad \Tilde{\varphi}||
  \end{split} \label{eq:1c}
\end{align}
\end{subequations}
The interpolation functions $h^{\text{fs}}(\varphi^\text{f}) = (\varphi^\text{f})^2(3-2\varphi^\text{f}) $ and $h^{\text{ff}}(\Tilde{\varphi}) = \tilde\varphi^2(3-2\tilde \varphi) $ for the fluid-solid and fluid-fluid interface are employed. For the phase variable of the fluid, a sinus profile of the form $\varphi^\text{f}(\eta) = 1/2\left[1-\sin{(\pi\eta/\delta_{\text{fs}})}\right]$ is applied, with $\delta_{\text{fs}}$ denoting the interface width and $\eta$ the normal coordinate pointing from the fluid to the solid at the fluid-solid-interface.  

\paragraph{Numerical Treatment}
This section briefly describes the algorithm for the phase-field equation~\eqref{eq:1c}. The approach of~\cite{reder2021phase,reder2022phase} is used to solve the Navier-Stokes equation. A fractional step method is used for the phase-field equation, which is represented as follows:
\begin{enumerate}
    \item First, we solve only the convective term, which gives
        \begin{align}
                    (\Tilde{\varphi^*}) = 
                    \intTime{t_n}{t_{n+1}}{ - \grad \Tilde{\varphi}^{n} \cdot \boldsymbol{u} }.
        \end{align}
     \item We then introduce the pseudotime  $\dimless{t}$ and then calculate the equilibrium according to
\begin{align}
\begin{split}
    \frac{\partial \Tilde{\varphi}^{**}}{\partial t^*} &= M \left[ \frac{36 \sigma}{\epsilon }(2\Tilde{\varphi^*}^3 -3\Tilde{\varphi^*}^2 + \Tilde{\varphi^*})    - \frac{\sigma \epsilon}{h^{\text{fs}}} \left( \grad \cdot h^{\text{fs}}   \grad \Tilde{\varphi^*} - h^{\text{fs}} ||\grad\Tilde{\varphi^*}||\grad \cdot \frac{\grad\Tilde{\varphi^*}}{||\grad\Tilde{\varphi^*}||} \right) \right. \\  &\qquad
    \left.- \frac{\partial_{\varphi^{\text{*}}}h^{\text{fs}}(\varphi^{*})||\grad\varphi^{*}||}{h^{\text{fs}}} (\sigma_{\text{2s}}-\sigma_{\text{1s}})\partial_{\Tilde{\varphi}}h^{\text{ff}}(\Tilde{\varphi^*})\right]
    \end{split}.
\end{align}
    \item In the last step, we apply the evaporation term
        \begin{align}
     \Tilde{\varphi}^{n+1} = \Tilde{\varphi}^{**} + \intTime{t_n}{t_{n+1}}{
     v^e ||\grad \Tilde{\varphi^{**}}||
     }.
\end{align}
\end{enumerate}
By splitting the phase-field equation, a more accurate representation of the equilibrium profile at the interface and the evaporation rate is achieved. Moreover,
the solution of the model equations is incorporated by applying the finite difference method on a Cartesian grid. The implementation was done using an in-house solver (PACE3D)~\cite{hotzer2018parallel}. 

\section{Validation}
\label{sec:validation}
The following chapter is divided into three main sections to discuss the validation of the model.  
Sec.~\ref{sec:dynamic-contanct-angle} examines the contact angle that occurs at a moving contact line between three phases.
Furthermore, in Sec.~\ref{sec:Rise}, the rise height of a liquid in a capillary is calculated and compared with the analytical solution, while in the final section (Sec.~\ref{sec:Channel}), the influence of different pore sizes on the drying process is investigated.
\subsection{Contact angle at a moving contact line}
\label{sec:dynamic-contanct-angle}
\begin{figure}[htbp] 
\centering
\begin{subfigure}[t]{.49\textwidth}
\centering
\includegraphics[scale =0.55]{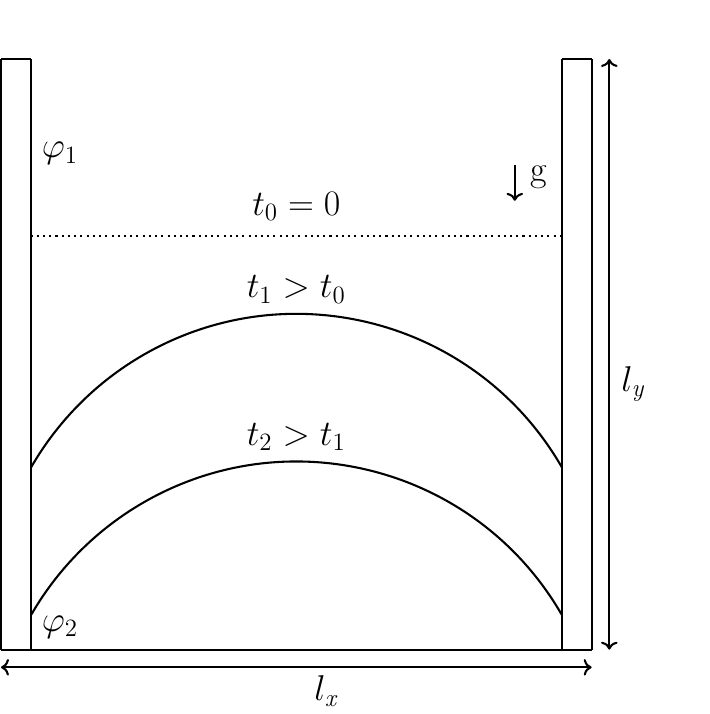}
    \subcaption{Schematic representation of the moving contact line problem.}\label{fig:Anfangsbedingung}
\end{subfigure} 
\begin{subfigure}[t]{.49\textwidth} 
\centering
\includegraphics[scale =1]{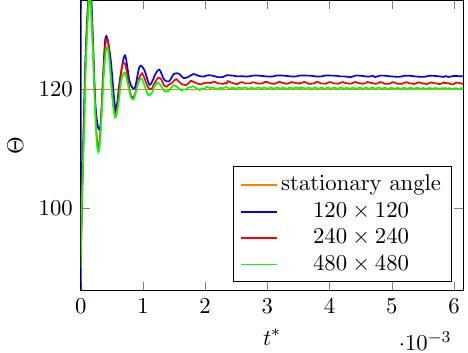}
\subcaption{Contact angle as a function of time for different grid refinements.}
\label{fig:delta-theta-t}
\end{subfigure} 
\caption{Validation case of an evaporating fluid in a capillary.}
\end{figure}

\noindent\textit{Setup and Parameters:} Fig.~\ref{fig:Anfangsbedingung} depicts the schematic representation of the first validation case.
The physical dimensions are $l_{\text{x}}=\SI{56.0}{\micro\metre}$ and  $l_{\text{y}} = \SI{56.0}{\micro\metre}$, with $l_{\text{x}}$ as the characteristic length $\charLength$.
A value of  $\sigma = \SI{7.3e-2}{\newton \per \metre}$ is set for the surface tension and a macroscopic equilibrium contact angle of $\theta=\ang{60}$ is employed. Furthermore, the gravitational body force is given by
$g=\SI{9.81}{\metre\per\square\second}$, whereas
for the fluids the following physical parameters apply. 
Fluid 1 $\varphi_1$ has a density of $\rho_1 = \SI{1.225}{\kg\per\cubic\metre}$ and a viscosity of $\mu_1 = \SI{1.72e-5}{\pascal\second}$, while fluid 2 $\varphi_2$ is defined by a density of $\rho_2 = \SI{997.0}{\kg\per\cubic\metre}$ and a viscosity of $\mu_2=\SI{1.0e-3}{\pascal\second}$. In addition, the factor $\kappa$ equals \SI{100}{\per\second} and the mobility is set to  $M=\SI{30.0}{\unitMobilityAC}$.
\\
\noindent\textit{Results:} Fig.~\ref{fig:delta-theta-t} shows the macroscopic contact angle $\Theta$, which is calculated according to the method given in Appendix~\ref{contact-angle-static}, as a function of a non-dimensionalized time $\dimless{t}=t\cdot\sqrt{g/\charLength}$, for different grid refinements and the stationary equilibrium contact angle ($\ang{180}-\theta$). 
The initially flat fluid interface becomes curved as the contact angle is established at the contact line where the solid and both fluid phases meet.
Oscillating movements of the interface are observed due to inertia effects, whereby the oscillations are damped over time.
After some time, the calculated contact angle settles to an average value~$\Theta_{\text{mean}}$, which shows only small fluctuations.
The mean contact angles $\ang{121.95}$, $\ang{120.85}$, and $\ang{120.12}$ are computationally measured for the resolutions $120\times120$, $240\times240$, and $480\times480$. 
It can be seen that the mean contact angle approaches the equilibrium angle with increasing resolution.
Additionally, the oscillation amplitude of $\Theta$ decreases with the refinement. It should be noted that the local contact angle is applied immediately according to equation~\eqref{eq:wettingBC}, while the macroscopic contact angle is the result of the surface shape (see Appendix~\ref{contact-angle-static}). This simulation study shows that the model is able to maintain the theoretical macroscopic contact angle in the three-phase region under dynamic conditions of a moving fluid interface. 
In addition, a convergence can be observed, which indicates that the numerical error decreases with higher resolution.   
\subsection{Rise of a fluid in a capillary}
\label{sec:Rise}
\noindent\textit{Setup and Parameters:} In this section, the capillarity between two fluids in a riser of width~$r$ is investigated. Regarding the filling height~$h$ in mechanical equilibrium, the analytical solution is given by equation
\begin{align}
     \hat{h} = \frac{h}{r} =\left(\frac{\sigma \cos{(\Theta)}}{\rho g } - A \right)\frac{1}{r^2},
     \label{eq:hfull}
\end{align}
with the  nondimensionalized height~$\hat{h}$, the surface tension~$\sigma$, the contact angle~$\Theta$, the density~$\rho$ of fluid 2, and the gravity~$g$. 
The derivation for the analytical solution and the definition of the area~$A$ can be found in the Appendix~\ref{sec:analytisch-steig}. 
Fig.~\ref{fig:Ausgangsaufbau-Riser} shows the initial setup.  
   \begin{figure}[htbp]
   \centering
       \includegraphics[scale =0.5]{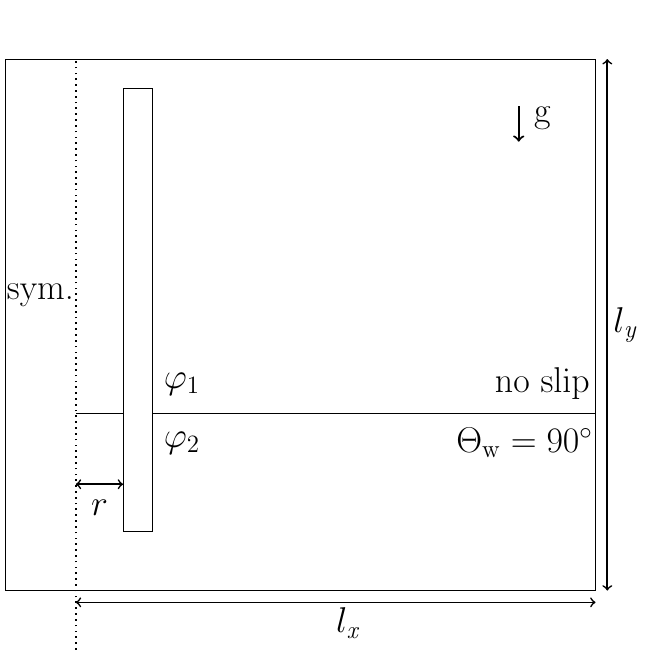}
       \caption[]{Initial setup for the rise of a fluid in a capillary.}
       \label{fig:Ausgangsaufbau-Riser}
   \end{figure} 
Due to the symmetry, only half of the riser is calculated, taking into account the physical dimensions of the system $l_{\text{x}} =\SI{10.0e-3}{\metre}$  and $l_{\text{y}} = \SI{55.0e-3}{\metre}$, and a riser  width of  $r = \SI{0.15e-3}{\metre}$. It is placed in a reservoir filled with water, represented by fluid 2 $\varphi_2$ with density $\rho_2 = \SI{997.0}{\kg\per\cubic\metre}$ and viscosity $\mu_{2}=\SI{1.0e-3}{\pascal\second}$. In addition, the surrounding fluid 1~$\varphi_1$ mimics air with a density of $\rho_1 = \SI{1.225}{\kg\per\cubic\metre}$ and a viscosity of $\mu_{1}= \SI{1.72e-5}{\pascal\second}$. The contact angle is given by $\theta = \ang{120}$, with a surface tension of $\sigma = \SI{7.3e-2}{\newton \per \metre}$. 
To validate the model, a grid refinement study is first carried out for a specified Bond number ($Bo =3.35\cdot 10^{-2} $),
which is defined as
\begin{align}
        Bo=\frac{\rho g \charLength^2}{\sigma},
\end{align}
with a characteristic length of~$\charLength=r$ for the present case.
While maintaining the physical domain size, simulations are performed with three different numerical resolutions, where the corresponding number of cells in the $x$- and $y$-directions is $100\times550$, $200\times1100$, and $400\times2200$. 
In addition, for the coarse resolution of $100\times550$ cells, the Bond number is varied by changing  the gravity $g$.
\\
\\
\noindent\textit{Results:} Fig.~\ref{fig:Result solution riser} shows the determined riser height $\hat{h}$ as a function of the Bond number $Bo$. In addition, the grid refinement study for a Bond number of $Bo =3.35\cdot 10^{-2} $ is depicted as an inlet picture. For both studies, the analytical solutions are marked by a brown line.
The simulation results for the grid refinement study are $\hat{h}=14.0$, $\hat{h}=14.27$ and $\hat{h}=14.33$ for the cell numbers $100 \times 550$, $200 \times 1100$ and $400 \times 2200$, respectively. 
Using the analytical solution $\hat{h}^{\text{anal}}=14.841$ the relative errors, defined by $\varepsilon = (\hat{h}^{\text{anal}}-\hat{h})/\hat{h}^{\text{anal}}$, are  $\varepsilon = \SI{5.56}{\percent}$, $\varepsilon = \SI{3.85}{\percent}$ and $\varepsilon = \SI{3.44}{\percent}$.
It follows that for a better grid refinement, the analytical solution is mapped more accurately, as the numerical error is reduced. 
This proves the convergence and successfully verifies the numerical solution. The simulated results for different Bond numbers for the cell numbers $n_{\text{x}} \times n_{\text{y}} = 100 \times 550$ are depicted as black dots in Fig.~\ref{fig:Result solution riser} illustrating that the model approximates the analytical solution well for all simulated Bond numbers.  
\begin{figure}[htbp]
\centering
\includegraphics[scale =1]{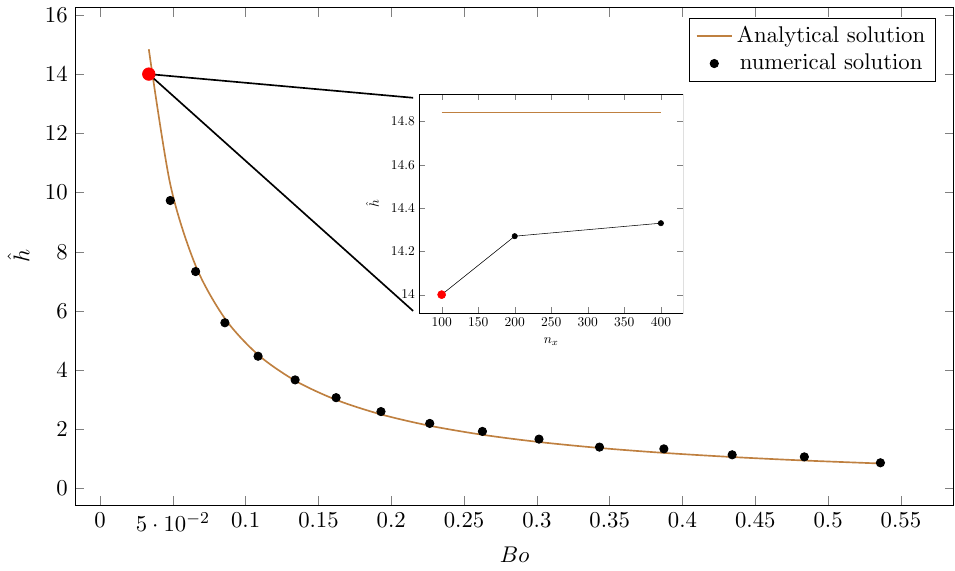}
       \caption{Rise of a liquid in a capillary: Dimensionless rise height $\hat{h}$ versus the Bond number for simulation results and the analytical solution.
Additionally, the effect of the grid refinement $n_{\text{x}}$ on the rise height $\hat{h}$ is shown as an inlet at a specific Bond number of $\mathit{Bo}=\num{0.05}$.}
       \label{fig:Result solution riser}
\end{figure}
\noindent

\subsection{Channel grid} 
\label{sec:Channel}
\begin{figure}[htbp]
\centering

\begin{subfigure}[c]{0.3\textwidth}
\centering
 \begin{tikzpicture}
 \node[anchor=south east] at (0,0){\setlength{\fboxsep}{0cm}
\setlength{\fboxrule}{0.03cm}
 \fbox{\includegraphics[scale=0.3]{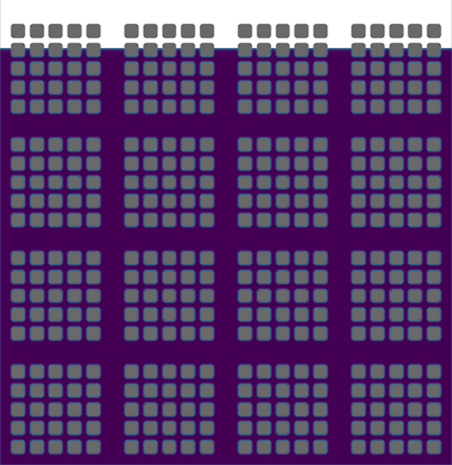}}};
\end{tikzpicture}
\end{subfigure} 
\begin{subfigure}[c]{0.3\textwidth}
\centering
 \begin{tikzpicture}
 \node[anchor=south east] at (0,0){\setlength{\fboxsep}{0cm}
\setlength{\fboxrule}{0.03cm}
 \fbox{\includegraphics[scale=0.3]{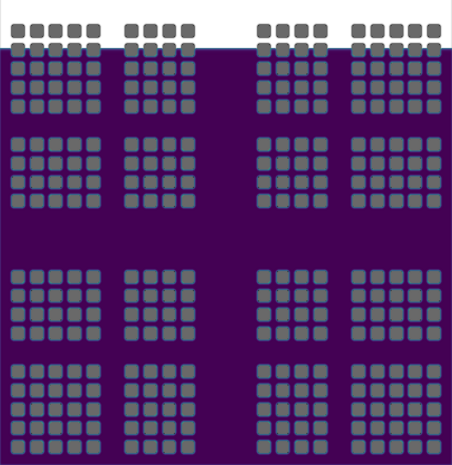}}};
\end{tikzpicture}
\end{subfigure} 
\begin{subfigure}[c]{0.3\textwidth}
\centering
 \begin{tikzpicture}
 \node[anchor=south east] at (0,0){\setlength{\fboxsep}{0cm}
\setlength{\fboxrule}{0.03cm}
 \fbox{\includegraphics[scale=0.3]{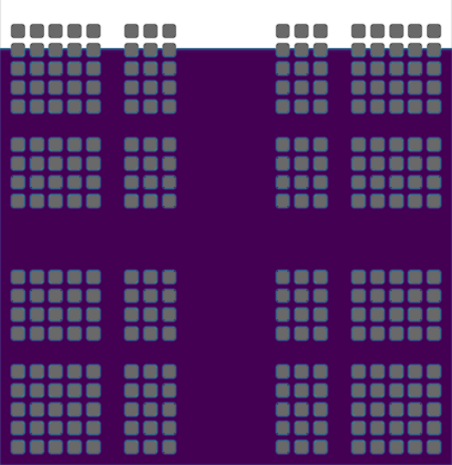}}};
\end{tikzpicture}
\end{subfigure} 
\begin{subfigure}[c]{0.30\textwidth}
\centering
 \begin{tikzpicture}
 \node[anchor=south east] at (0,0){\setlength{\fboxsep}{0cm}
\setlength{\fboxrule}{0.03cm}
 \fbox{\includegraphics[scale=0.3]{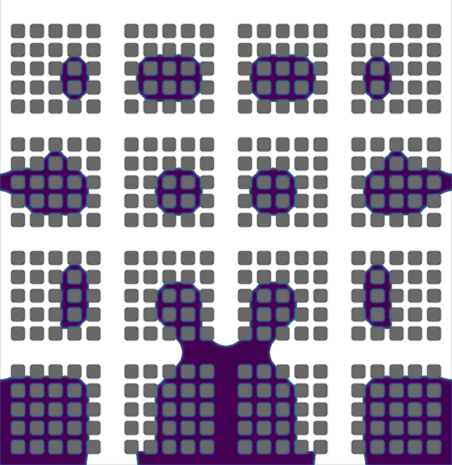}}};
\end{tikzpicture}
\end{subfigure} 
\begin{subfigure}[c]{0.3\textwidth}
\centering
 \begin{tikzpicture}
 \node[anchor=south east] at (0,0){\setlength{\fboxsep}{0cm}
\setlength{\fboxrule}{0.03cm}
 \fbox{\includegraphics[scale=0.3]{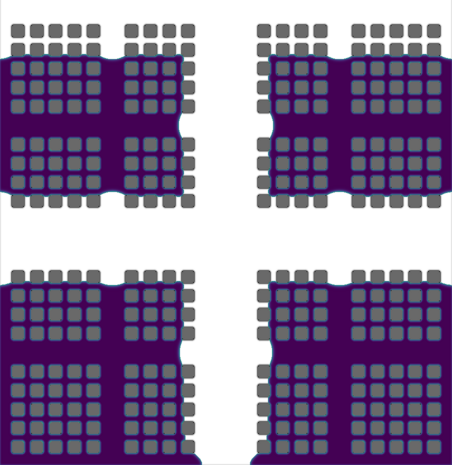}}};
\end{tikzpicture}
\end{subfigure} 
\begin{subfigure}[c]{0.3\textwidth}
\centering
 \begin{tikzpicture}
 \node[anchor=south east] at (0,0){\setlength{\fboxsep}{0cm}
\setlength{\fboxrule}{0.03cm}
 \fbox{\includegraphics[scale=0.3]{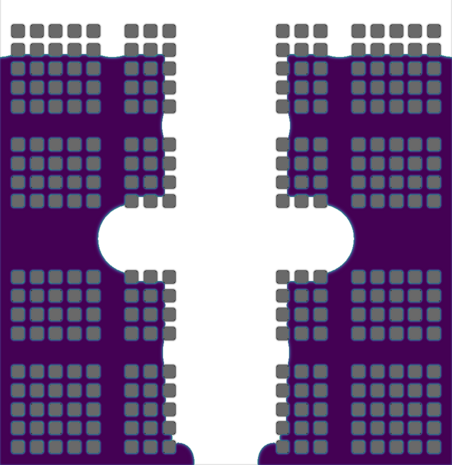}}};
\end{tikzpicture}
\end{subfigure} 
\caption{Simulation results for different pore structures, with the initial condition in the top row and time of breakthrough in the bottom row. The fluids, referred to as Fluid 1 (air) and Fluid 2 (liquid), are shown in white and blue, respectively. The solid phase is  displayed in gray. }
\label{fig:channelgrid}
\end{figure}

In this section, the influence of different pore sizes on pore emptying during drying is analyzed.
The simulation parameters are identical to those from Sec.~\ref{sec:dynamic-contanct-angle}, with the exception of the contact angle, which for this study is $\theta = \SI{90}{\degree}$.
A resolution of  $480 \times 480$  with a physical domain size of $l_{\text{x}}=\SI{56.0}{\micro\metre}$ and $l_{\text{y}} = \SI{56.0}{\micro\metre}$ is used.
Three different configurations, which are shown in Fig.~\ref{fig:channelgrid}, are considered. 
These configurations differ in the pores that are located horizontally and vertically in the center of the simulation area.
Starting from the reference configuration, which is depicted on the left in Fig.~\ref{fig:channelgrid}, the pores located in the horizontal and vertical center are enlarged for the second configuration (center) by removing rectangles.
For the third configuration (right), the vertical pore is enlarged by further removing rectangles. 
This approach allows the simulation of different pore size distributions and the subsequent analysis of their effects during the drying process, in analogy to the setup used in
~\cite[Fig. 4.17]{kharaghani2020drying}. 
The bottom row of Fig.~\ref{fig:channelgrid} shows the different configurations at the time of breakthrough, which indicates the point at which the air reaches the bottom of the system.
For the reference configuration, it can be observed that the liquid evaporates mainly into the larger pores, but also into the smallest pores. 
In the second configuration, which has larger pores in the horizontal and vertical center, the liquid evaporates preferentially into these pores. 
Additionally, the liquid begins to evaporate in the next smaller pores, while the smallest pores are still filled in this configuration. 
For the last configuration, the result is similar to the other configurations, with the larger pores emptying first, followed by the smaller pores. 
In contrast to the other configurations, pinning of the liquid at the original fill level can be observed. 
In summary, it can be said that large capillaries empty first before smaller capillaries, which is consistent with the findings from the literature~\cite{metzger2005influence}. 
The possibility of pinning on the surface is given, which has also been observed in the drying of battery electrodes~\cite{jaiser2017microstructure}, and can be described analytically for a two-capillary system~\cite{krischer2013wissenschaftlichen}. 
This pinning is reflected in a height difference between a continuous line connecting the highest and lowest positions of the liquid and will be used as a parameter for the subsequent analysis.  
Wrapping up, the result of this simulation study shows the expected behavior indicating that the present model is suitable to predict pore emptying processes.
\section{Application of the model to the drying of hard carbon electrodes for batteries }
\label{sec:application}
In this chapter, the drying of the solvent of battery electrodes is simulated using the proposed model. Herein, the drying is modeled as a convective capillary flow, while transport phenomena of the evaporated liquid in the gas phase are excluded. 
The influence of viscosity on the drying process is investigated for two different microstructures. 
In addition, the effect of different contact angles on pore emptying is investigated for one representative microstructure.
\subsection{Preprocessing}
\begin{figure}[htbp]
\centering
\begin{subfigure}[c]{0.49\textwidth}
\centering
\vspace{-0.1cm}
\begin{tikzpicture}
 \node[anchor=south east] at (0,0){
\includegraphics[scale=0.11,clip, trim = 0 144 0 0]{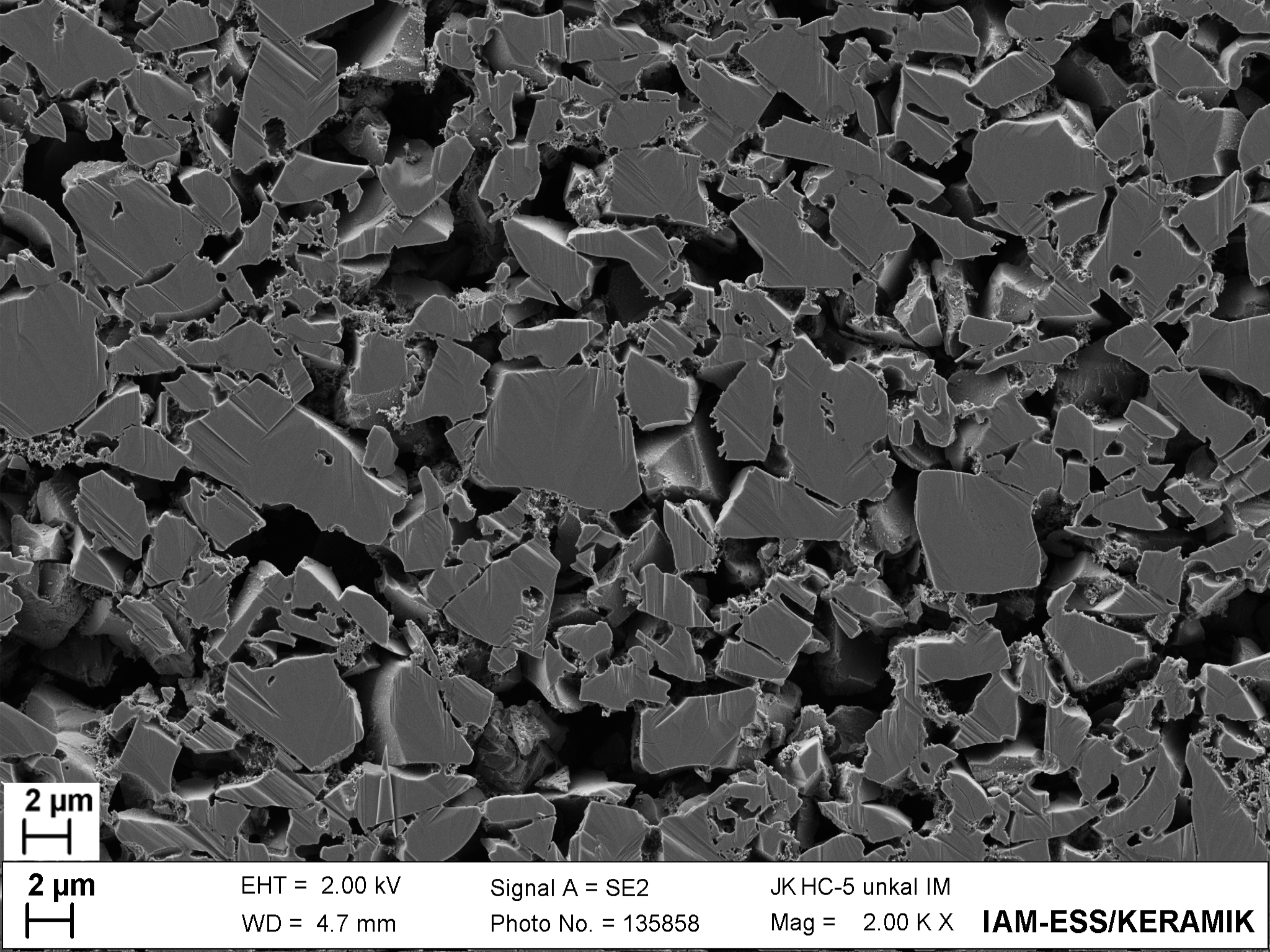} };
\end{tikzpicture}
\subcaption{SEM image of the microstructure  HC-B.}
 \label{fig:HC-B-mikro}
\end{subfigure} 
\begin{subfigure}[c]{0.49\textwidth}
\centering
\begin{tikzpicture}
 \node[anchor=south east] at (0,0){
\setlength{\fboxsep}{0cm}
\setlength{\fboxrule}{0.03cm}
\centering
\fbox{\includegraphics[scale=0.355]{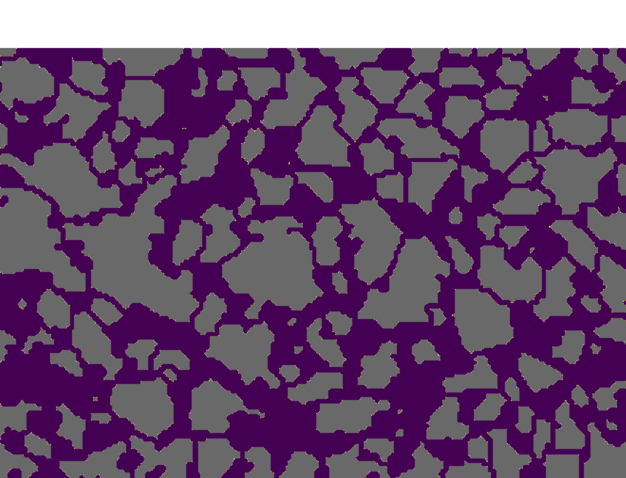}}};
\node[anchor=south east] at (-0.1,5.7){\small air}; 
\end{tikzpicture}
\subcaption{Initial simulation input for the microstructure HC-B.}
 \label{fig:HC-B-sim}
\end{subfigure} 
\begin{subfigure}[c]{0.49\textwidth}
\centering
\vspace{-0.1cm}
\begin{tikzpicture}
 \node[anchor=south east] at (0,0){
\includegraphics[scale=0.11,clip, trim = 0 144 0 0]{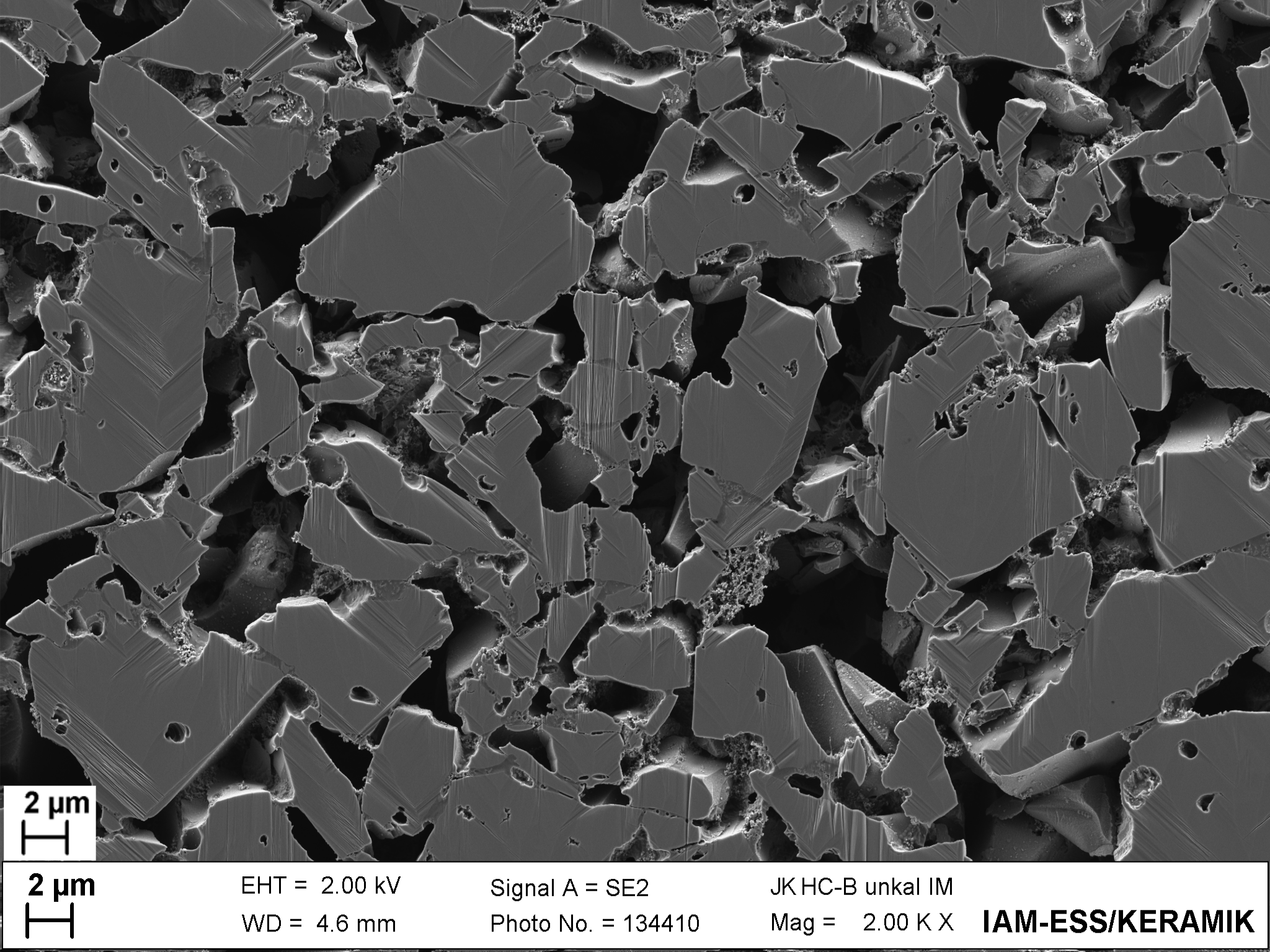}};
\end{tikzpicture}
\subcaption{SEM image of the microstructure HC-C.}\label{fig:HC-C-mikro}
\end{subfigure} 
\begin{subfigure}[c]{0.49\textwidth}
\begin{tikzpicture}
 \node[anchor=south east] at (0,0){
\setlength{\fboxsep}{0cm}
\setlength{\fboxrule}{0.03cm}
\centering
\fbox{\includegraphics[scale=0.355]{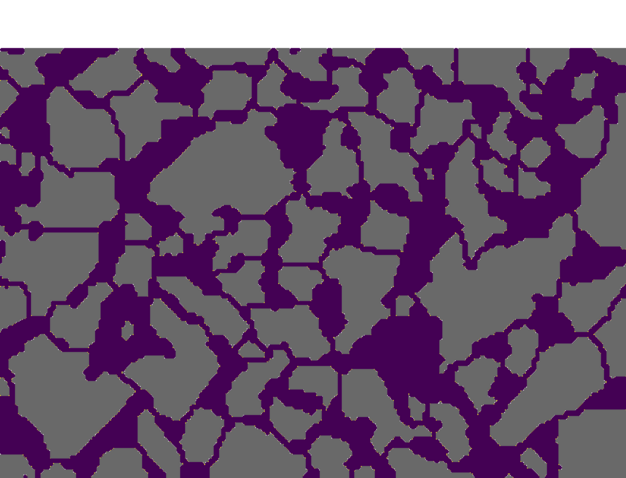}}} ;
\node[anchor=south east] at (-0.1,5.7){\small air}; 
\end{tikzpicture}
\subcaption{Initial simulation input for the microstructure HC-C.}
\end{subfigure} 
\caption{Microstructures of hard carbon~\cite{klemens2023process} with different mean particle diameters and pore size distributions as well as the corresponding initial microstructures for the simulations.}
\label{fig:microstructures}
\end{figure}

\paragraph{Microstructures} The simulations are performed on the basis of two images of dried hard carbon electrodes. 
These are integrated into the simulation environment, with the help of a Kadi4Studio workflow~\cite{griem2022kadistudio} and an imageJ macro. 
Fig.~\ref{fig:microstructures} shows the two different microstructures from Klemens et al.~\cite{klemens2023process} with the corresponding microstructures as input for the simulations. 
For a better readability, the names HC-B and HC-C are adapted accordingly.
Microstructure HC-B has a smaller average particle diameter and a less broad pore and particle size distribution than the microstructure HC-C.
Furthermore, the microstructure HC-B is defined by a specific area weight of $M_{\text{S}}= \SI{78.4e-3}{\kg\per\meter\squared}$ and a volume fraction of $\chi_{\text{solid}}=0.246$, while the microstructure HC-C  has the values $M_{\text{S}}= \SI{79.5e-3}{\kg\per\meter\squared}$ and $\chi_{\text{solid}} =0.245$.
The physical lengths of the microstructures are $l_{\text{x}}=\SI{55.0}{\micro\metre}$ and $l_{\text{y}}=\SI{37.5}{\micro\metre}$. 
In order to take the initial air layer into account, the computational domain is extended in preprocessing, considering a domain height of $l_{\text{y}}=\SI{42}{\micro\metre}$.
\\
\\
\paragraph{Parameters} The material parameters for the solvent are selected with $\rho_{\text{solvent}}= \SI{997.0}{\kg\per\cubic\metre}$ for the density and $\mu_{\text{solvent}}= \SI{1.0}{\pascal\second}$ for the viscosity. It should be noted that due to the thickener carboxymethyl cellulose (CMC), the viscosity differs from the viscosity of water and is chosen according to the literature~\cite{benchabane2008rheological}. 
Both the fluid and the battery slurry are shear-thinning~\cite{klemens2023process}, which can be modeled with a rate-dependent viscosity and thus a generalized Newtonian behavior. 
Subsequently, in the present model all fluids are considered as a Newtonian fluid for sake of simplicity.
Regarding the air phase, a density of $\rho_{\text{air}}= \SI{1.225}{\kg\per\cubic\metre}$ and a viscosity of $\mu_{\text{air}}= \SI{1.72e-5}{\pascal\second}$ are assumed.
As the density interpolation becomes unstable at high density ratios (cf. \cite{Abels2012}), the air density is artificially increased by a factor of 4 for the subsequent simulations.
A surface tension between the different fluids of $\sigma = \SI{7.3e-2}{\newton\per\metre}$  with a contact angle of \ang{90} is applied. Furthermore, the gravity is set to $\gravity=\SI{9.81}{\metre\per\square\second}$. 
The mobility is defined by $M=\SI{30.0}{\unitMobilityAC}$, whereby an area-specific evaporation rate of $\Dot{r}_{\text{solvent}}$ = \SI{9}{\g\per\meter\squared\per\second} is used, which is chosen relatively high in the context of film drying~\cite{klemens2023challenges} to avoid excessive simulation times.
All physical parameters and the corresponding dimensionless values can be found in Table~\ref{tab:Ausgangsparameter-HC}.
For the cell numbers the values $n_x=\num{592}$ and $n_y=\num{448}$ are used, leading to the dimensionless spatial step sizes of $\dimless{\Delta x} := \Delta x/\chr{l} = 2.78716\cdot 10^{-5}$ and $\dimless{\Delta y} := \Delta y/\chr{l}= 2.8125\cdot 10^{-5}$, respectively.
\begin{table}[htbp]
\centering
    \captionof{table}{Overview of the simulation parameter for hard carbon microstructures.}
\begin{ruledtabular}
  \begin{tabular}{l c c c c c }
    parameter & symbol &  physic. value & unit & nondim value\\
    \hline  
    physical length & $l_{\text{x}}$ & $55.0$ & \si{\micro\meter} &$0.0165$\\
    physical height & $l_{\text{y}}$ & $42.0$& \si{\micro\meter} &$0.0126$\\
    density hard carbon & $\rho_{\text{HC}}$ & $2069.0$ & \si{\kg\per\cubic\metre} & $2.08$ \\
    density  solvent & $\rho_{\text{solvent}}$ & $997.0$ & \si{\kg\per\cubic\metre} & $1.0$ \\
    density air & $\rho_{\text{air}}$ & $1.225$ & \si{\kg\per\cubic\metre} & $1.23\cdot 10^{-3}$\\
    viscosity solvent & $\mu_{\text{solvent}}$ & $1.0$ & \si{\pascal\second}&$5.26\cdot 10^{-3}$ \\
    viscosity air &$\mu_{\text{air}}$ & $1.72\cdot 10^{-5}$ & \si{\pascal\second}&$9.05\cdot 10^{-5}$\\
    surface tension & $\sigma$ & $7.3\cdot 10^{-2}$ & \si{\newton\per\metre} &$6.72$ \\
    gravity & $g$ & $9.81$ & \si{\metre\per\square\second} & $10.0$\\
    mobility & $M$ & $30$ & \si{\unitMobilityAC} & $1.81$\\
    \end{tabular}
    \end{ruledtabular}
    \label{tab:Ausgangsparameter-HC}
\end{table}
\\
\\
\paragraph{Dimensional analysis}
Using the parameters from Table~\ref{tab:Ausgangsparameter-HC}, a non-dimensionalization analysis is performed. In addition, the characteristic velocity is described by $\charVelocity =\sqrt{\gravity\cdot \charLength}=\SI{0.023}{\metre\per\second}$, with the characteristic length scale $\charLength=l_{\text{x}}=\SI{55.0}{\micro\meter}$. 
This results in the relevant dimensionless variables $\Bo_{\text{s}}=\num{4.5e-4}$, $\Ca_{\text{s}}=\num{0.35}$, $\Reynolds_{\text{s}}=\num{1.27e-3}$, $\Bo_{\text{a}}=\num{5.51e-7}$, $\Ca_{\text{a}}=\num{6.05e-6}$ and $\Reynolds_{\text{a}}=\num{0.091}$ for the solvent and air. 
\\
\\
To enable short simulation times, we use the approach of Wolf et al.~\cite{wolf2024computational} and scale the physical problem.
For this, the drying rate is corrected upwards by the factor $\num{5.0e4}$, while the viscosity of the two fluids present is scaled downwards by the same factor.
This results in a constant capillary number, which represents the ratio of viscous forces to capillary forces and is defined later.
\\
\\
Various parameters are used to evaluate the simulation results. 
These are controlled by pore breakthroughs, which have also been observed in experiments~\cite{kumberg2021influence} and are defined by the breakthrough of the air onto the substrate. 
First, the non-dimensionalized time ($t^{\text{breakthrough}}$) of the breakthrough is evaluated.
The physical time can be calculated according to $t^{\text{physical}} = 50000 \cdot t\cdot t^{\text{char}} $, with $50000$ as the scaling factor of the evaporation rate, $t^{\text{char}} = \sqrt{l_{\text{char}}/g}$ as the characteristic time, and $t$ as the non-dimensionalized time. 
Thereby achieving, a drying time of approximately \SI{4.81}{\second} for the second drying stage, with an evaporation rate of \SI{9}{\g\per\meter\squared\per\second}, which can be achieved experimentally~\cite{klemens2023process}.
Furthermore, the height difference, denoted as $\Delta h$, is examined at the breakthrough time $(t^{\text{breakthrough}})$, as shown in the inlet pictures in Fig.~\ref{fig:norm-delta-h-vs-Ca}.
The height difference is defined as the maximum difference between $y$-coordinates of the solvent front, which are continuously connected to each other and thus enable the capillary transport of solvent and binder. 
For further evaluation, the height difference is normalized to the initial film height $h_{\text{film}}^{t=0}$ and the corresponding abbreviation $ \auswertungDeltah= \Delta h/h_{\text{film}}^{t=0}$ is introduced.
Another parameter that is examined is the normalized $\hat{x}$-coordinate, which specifies the $x$-coordinate of the breakthrough normalized by $l_{\text{x}}$. 
Simulations results are evaluated in relation to the capillary number
\begin{align}
    Ca = \frac{\mu_{\text{solvent}} u_{\text{char}}}{\sigma}.
\end{align}
\subsection{Variation of the capillary number}
\begin{figure}[htbp]
\centering
\begin{subfigure}[c]{1\textwidth}
\includegraphics[scale =1]{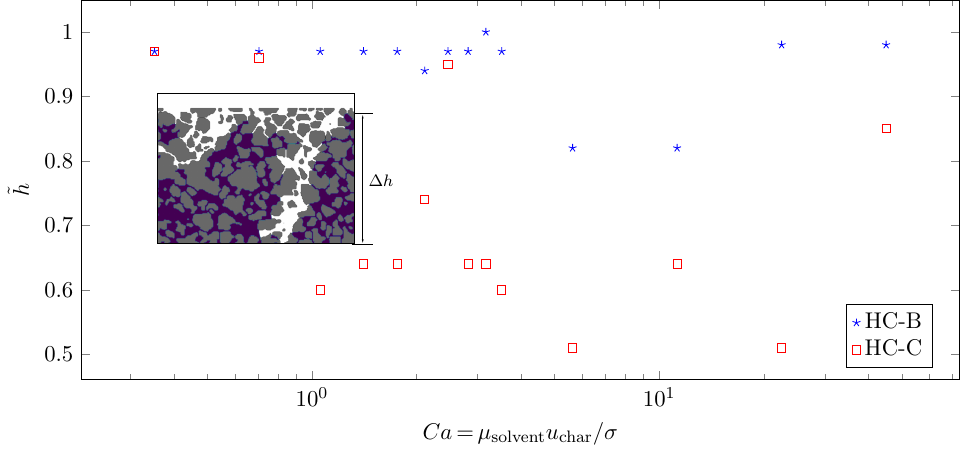}
\subcaption{Simulated normalized height difference $\auswertungDeltah  (=\Delta h/h_{\text{film}}^{t=0})$ of the film, as a function of the capillary number $Ca$, for two microstructures. The blue and red symbols refer to the microstructures HC-B and HC-C, respectively.}
\label{fig:norm-delta-h-vs-Ca}
\end{subfigure}
\begin{subfigure}[c]{0.49\textwidth}
\setlength{\fboxsep}{0cm}
\setlength{\fboxrule}{0.03cm}
\centering
\fbox{\includegraphics[scale=0.3]{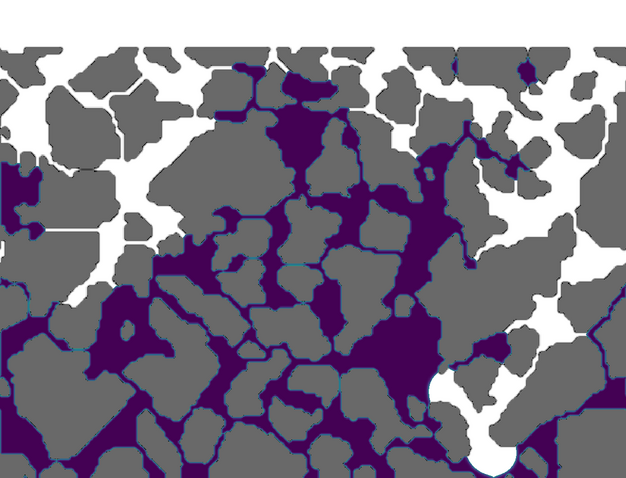}}  
\subcaption{HC-C $(Ca=0.352)$ at $t^{\text{breakthrough}} = 0.00099$. }
\label{fig:HC-C-x3-Ca-0.352}
\end{subfigure}
\begin{subfigure}[c]{0.49\textwidth}
\centering
\setlength{\fboxsep}{0cm}
\setlength{\fboxrule}{0.03cm}
\centering
\fbox{\includegraphics[scale=0.3]{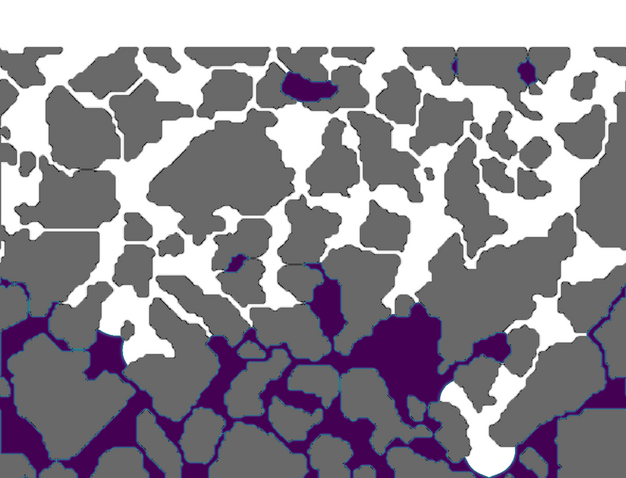}}  
\subcaption{HC-C $(Ca=5.631)$ at $t^{\text{breakthrough}} = 0.00161 $.}
\label{fig:HC-C-x3-Ca-5.631}
\end{subfigure}
\caption{Dependence of the normalized height difference $\auswertungDeltah$ of the breakthrough point on the capillary number $Ca$ (top). Microstructure images of simulations for selected combinations of capillary number and breakthrough time (bottom).}
\end{figure}

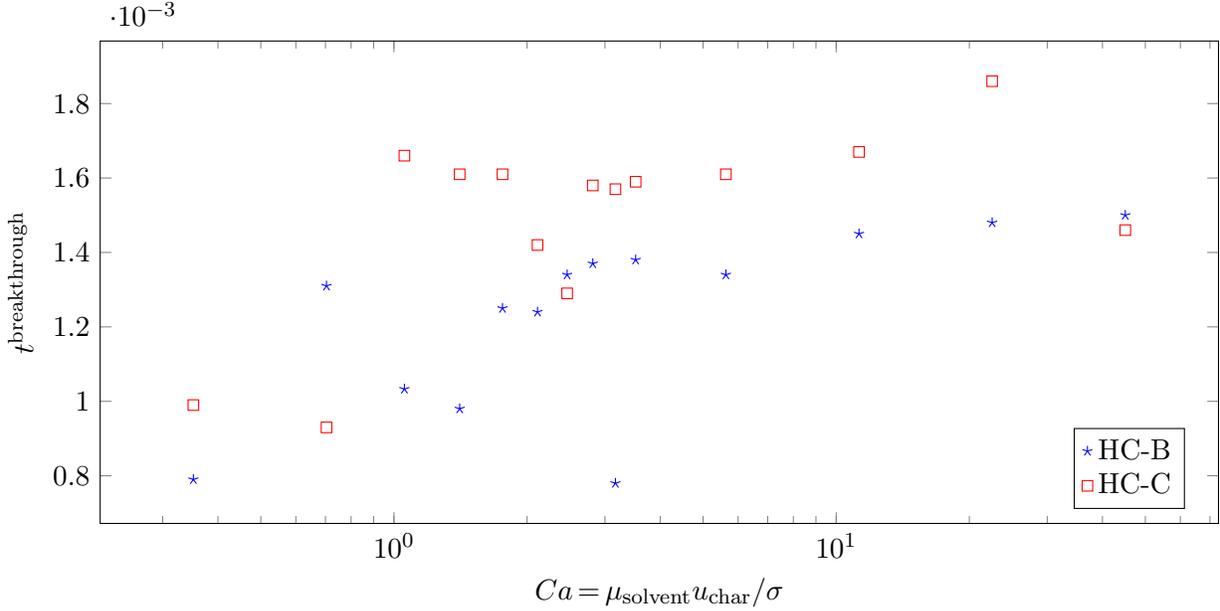
\begin{figure}[htbp]
\centering
\centering
\begin{tikzpicture}
\begin{axis}[
  width=\textwidth,
  height=8cm,
  ylabel=$t^{\text{breakthrough}}$,
  xmode=log,
  xlabel = $Ca \, {=} \,\mu_{\text{solvent}} u_{\text{char}}/\sigma$,
  cycle list name = varyPaired,
  legend pos=south east
]
 \addplot[
        only marks,
        mark=star,
        color=blue,
    ] coordinates {
        (0.352,0.00079)
        (0.704,0.00131)
        (1.056,0.001033)
        (1.408,0.00098)
        (2.464,0.00134)
        (2.816, 0.00137)
        (3.168,0.00078)
        (3.52,0.00138)
        (11.262,0.00145)
        (22.524,0.00148)
        (45.047,0.0015)
        (1.76,0.00125)
        (2.112,0.00124)
        (5.631,0.00134)
    };
\addlegendentryexpanded{HC-B}
 \addplot[
        only marks,
        mark=square,
        color=red,
    ] coordinates {
        (0.352,0.00099)
        (1.056,0.00166)
        (1.408,0.00161)
        (1.76,0.00161)
        (2.112,0.00142)
        (2.464,0.00129)
        (2.816,0.00158)
        (3.168,0.00157)
        (5.631,0.00161)
        (11.262,0.00167)
        (22.524,0.00186)
        (0.704,0.00093)
        (45.047,0.00146)
        (3.52,0.00159)
    };
\addlegendentryexpanded{HC-C}
\end{axis}
\end{tikzpicture}
\caption{Breakthrough time $t^{\text{breakthrough}}$ as a function of the capillary number $Ca$. The microstuctures HC-B and HC-C are shown in blue and red, respectively.}
\label{fig:t-durchbruch-vs-Ca}
\end{figure}

\begin{figure}
    \centering
\includegraphics[scale =1]{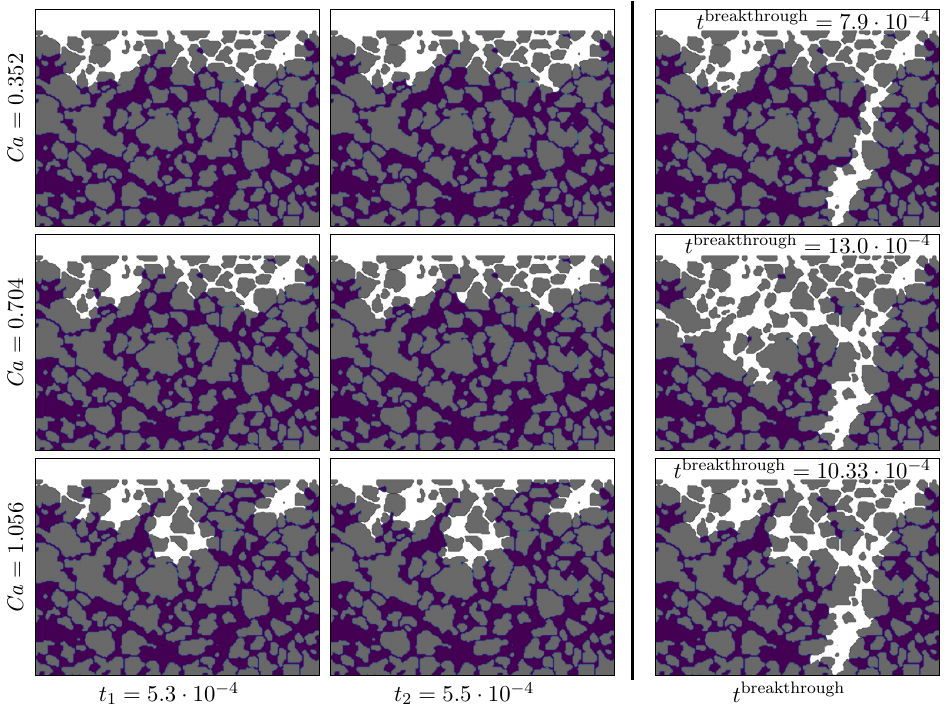}
    \caption{Temporal evolution of the microstructure HC-B for different capillary numbers $Ca$, at fixed times $t_1=\num{5.3e-4}$, $t_2=\num{5.5e-4}$ and measured breakthrough times: $t^{\text{breakthrough}}=\num{7.9e-4}$, $t^{\text{breakthrough}}=\num{13.33e-4}$ and $t^{\text{breakthrough}}=\num{10.33e-4}$ for  $\Ca=\num{0.352}$, $\Ca=\num{0.704}$ and $\Ca=\num{1.056}$. The air and solvent particles are shown in white, blue, and gray, respectively.}
    \label{fig:time-evolution}
\end{figure}

\paragraph{Height difference of the liquid front}
Fig.~\ref{fig:norm-delta-h-vs-Ca} shows the normalized difference between the maximum height and the minimum height at the time of breakthrough onto the substrate as a function of the capillary number. 
For the microstructure HC-B the results are displayed in blue, while the results for HC-C are plotted in red. 
In the case of the microstructure HC-B, it can be seen that for most cases, with the exception of the capillary numbers $\Ca=\num{5.631}$ and $\num{11.262}$, the normalized height difference is approximately $\auswertungDeltah= 0.97$. 
So there are still places where a solvent column reaches the film height, whereas otherwise columns have fallen dry depending on their shape and diameter.
For the capillary numbers $\Ca=\num{5.631}$ and $\num{11.262}$, the normalized height difference is $ \auswertungDeltah= 0.82$. 
The behavior for the microstructure HC-C differs from that of the other microstructure. 
In general, the scatter of the values for $ \auswertungDeltah$ is higher than for the microstructure HC-B. 
At a capillary number of $\Ca=\num{1.056}$, the value for the normalized height difference drops to approximately $ \auswertungDeltah= 0.64$ and scatters around this value. 
An exception to this are the capillary numbers $\Ca=\num{2.464}$ and $\Ca=\num{45.047}$ with the values $ \auswertungDeltah= \num{0.94}$ and $ \auswertungDeltah= \num{0.85}$, respectively. 
Figs.~\ref{fig:HC-C-x3-Ca-0.352} and \ref{fig:HC-C-x3-Ca-5.631} show microstructure images of the solvent distribution at the time of breakthrough for the capillary numbers $\Ca=\num{0.352}$ and $\Ca=\num{5.631}$ of the microstructure HC-C. 
Comparing the two snapshots, a different behavior can be observed. 
In Fig.~\ref{fig:HC-C-x3-Ca-0.352}, the liquid front pins much closer to the microstructure surface than in Fig.~\ref{fig:HC-C-x3-Ca-5.631}. 
Similar pinning on the film surface is also reported in other simulation studies~\cite{metzger2007isothermal} as well as in experiments~\cite{kumberg2019drying, Klemens2022} and is based on the capillary forces acting in a pore system. 
Whether pinning occurs depends on the viscosity and the pore size distribution of the microstructure~\cite{metzger2007isothermal}. 
While capillary forces transport the liquid to the surface during drying, this effect is counteracted by the damping forces of viscosity. 
Therefore, the difference in the behavior observed in Fig.~\ref{fig:norm-delta-h-vs-Ca} can be explained by the dependence on the capillary number. 
If the capillary number increases due to a higher viscosity of the liquid, the damping forces increase. 
Thus, the probability of transport to the surface decreases. 
In general, the solvent in the microstructure HC-B is more likely to pin to the top of the film, compared to the HC-C microstructure, due to the narrow pore size distribution, which is also reported by Metzger et al.~\cite{metzger2005influence,metzger2007influence}. 
Moreover, the microstructure HC-C shows a wider scatter for the normalized height as a result of a wider pore size distribution that allows for more capillary veins. 
\\
\\
\paragraph{Breakthrough time}
Fig.~\ref{fig:t-durchbruch-vs-Ca} displays the time of breakthrough as a function of the capillary number for the two different microstructures HC-B and HC-C, with blue and red symbols. 
A similar relationship between the capillary number, the viscosity, and the breakthrough time is found for both microstructures. 
In the case of the microstructure HC-C, the breakthrough times are greater than for the microstructure HC-B for all cases, except for the capillary numbers $\Ca=\num{0.704}$, $\Ca=\num{2.464}$, and $\Ca=\num{45.047}$. 
For the microstructure HC-C, the breakthrough time for the capillary numbers $\Ca=\num{2.112}$, $\Ca=\num{2.464}$, and $\Ca=\num{45.049}$ is noticeable in terms of a shorter breakthrough time, compared to the surrounding capillary numbers of the microstructure HC-C. 
A remarkable result is the observation that the breakthrough time increases with larger capillary numbers, since a higher viscosity dampens the speed of the capillary flow in the solvent. 
This reduces the probability that a specific capillary will form and break through to the substrate. 
In the limiting case of $\Ca\to\infty$, a horizontal liquid front is expected during evaporation. 
However, the breakthrough time is not only influenced by the speed in the capillary, but also by the length of the breakthrough path that is taken. 
It can be observed that, depending on the capillary number, different breakthrough paths are preferred, which is indicated by different $x$-coordinates of the breakthrough point, at which the air reaches the substrate first (see Appendix~\ref{coordinate}).
Additionally, the height reduction within a certain capillary of the microstructure can be temporarily stopped, depending on the capillary number, and influence the breakthrough time. 
This illustrates the influence of the microstructure and its particle distribution. 
A more detailed discussion of the location of the various breakthrough points can be found in the appendix~\ref{coordinate}.
\\
\\
\paragraph{Temporal development} In order to explain some of the deviations in the results, the temporal development of the microstructure HC-B for the capillary numbers $\Ca=\num{0.352}$, $\Ca=\num{0.704}$, and $\Ca=\num{1.056}$ is examined as an example. 
Fig.~\ref{fig:time-evolution} includes the simulation results for three different points in time, labeled $t_1$, $t_2$, and $t^{\text{breakthrough}}$.
The fluids are visualized in white (air) and blue (solvent), while the particles are shown in gray.
For the capillary number $\Ca=\num{0.704}$, the breakthrough time is significantly higher than for the capillary numbers $\Ca=\num{0.352}$ and $\Ca=\num{1.056}$.
This contradicts the trend in Fig.~\ref{fig:x-durchbruch-vs-Ca}, where the time increases for higher capillary numbers. 
At time $t_1$, the pore emptying is identical for the capillary numbers $\Ca=\num{0.352}$ and $\Ca=\num{0.704}$. 
For $\Ca=\num{1.056}$, the picture is already different at this point in time. 
At time $t_2$, the capillary numbers $\Ca=\num{0.352}$ and $\Ca=\num{0.704}$ differ. 
For a value of $\Ca=\num{0.704}$, a temporally limited pinning of the solvent in the capillary occurs in the right-hand area of the microstructure, which is not observed at a capillary number of $\Ca=\num{0.352}$. 
The effect of this pinning can be seen at the breakthrough time $t^{\text{breakthrough}}$. 
With the lower capillary number, the liquid front has not penetrated as far into the microstructure, which is associated with an earlier breakthrough time. 
A comprehensive understanding of such behavior is only possible  due to the information provided by the developed model across the entire field.
\\
\\
\paragraph{Summary} In this section, pore emptying is investigated under a number of boundary conditions. 
These boundary conditions are the viscosity and the property of the microstructure, which is not only considered by static parameters. 
It is shown that the effective capillary leading to breakthrough depends on the capillary number and the microstructure properties. 
Here, the effective capillary refers to the path of the liquid front up to the breakthrough. 
The height difference of the liquid front, which describes the pinning on the surface of a pore structure, is also dependent on these parameters. 
Outliers in the simulation results could be attributed to other effective capillaries found in the microstructure. 
The investigation of this behavior is only possible because a model approach is used that provides full-field information on various quantities. 
This includes the phase-field variable, which enables a precise representation of the solvent over time. 
In summary, computational modeling provides novel insights into the microstructural properties and developments during the drying process of battery electrodes. 

\subsection{Variation of the contact angle}
\begin{figure}[htbp]
\centering
\includegraphics[scale =1]{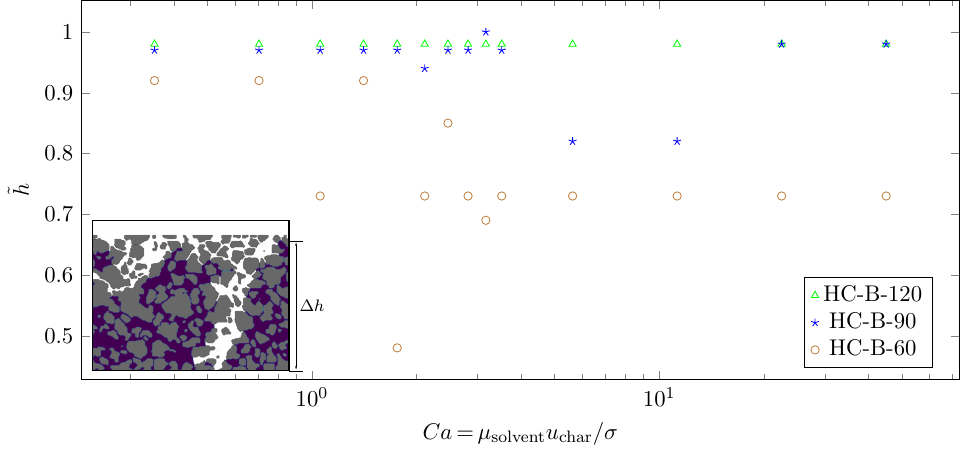}
\caption{Dependence of the normalized film height $ \auswertungDeltah$ on the capillary number for different contact angles at the time of breakthrough. The contact angles \ang{60}, \ang{90}, and \ang{120} are shown in brown, blue, and green.}
\label{fig:deltah-vs-ca-contact}
\end{figure}

\begin{figure}[htbp]
\centering
\centering
\begin{tikzpicture}
\begin{axis}[
  width=\textwidth,
  height=8cm,
  ylabel=$t^{\text{breakthrough}}$,
  xmode=log,
  xlabel = $Ca \, {=} \,\mu_{\text{solvent}} u_{\text{char}}/\sigma$,
  cycle list name = varyPaired,
  legend pos=south east
]
\addplot[
        only marks,
        mark=triangle,
        color=green,
    ] coordinates {
        (0.352,0.00086)
        (0.704,0.00077)
        (1.056,0.00088)
        (1.408,0.00089)
        (1.76,0.00080)
        (2.112,0.00079)
        (2.464,0.00079)
        (2.816,0.00079)
        (3.168,0.00079)
        (3.52,0.00079)
        (5.631, 0.00090)
        (11.262,0.00089)
        (22.524,0.00092)
        (45.047,0.00134)       
    };
\addlegendentryexpanded{HC-B-120}
 \addplot[
        only marks,
        mark=star,
        color=blue,
    ] coordinates {
        (0.352,0.00079)
        (0.704,0.00131)
        (1.056,0.00103)
        (1.408,0.00098)
        (2.464,0.00134)
        (2.816, 0.00137)
        (3.168,0.00078)
        (3.52,0.00138)
        (11.262,0.00145)
        (22.524,0.00148)
        (45.047,0.0015)
        (1.76,0.00125)
        (2.112,0.00124)
        (5.631,0.00134)
    };
\addlegendentryexpanded{HC-B-90}
\addplot[
        only marks,
        mark=o,
        color=brown,
    ] coordinates {
        (0.352,0.00142)
        (0.704,0.00128)
        (1.056,0.00195)
        (1.408,0.00141)
        (1.76,0.00228)
        (2.464,0.00145)    
        (3.168,0.00186)    
        (5.631, 0.00188)   
        (2.112,0.00192)
        (22.524,0.00187)
        (45.047,0.00203)  
        (11.262,0.00207)  
        (2.816,0.00185)
        (3.52,0.00197) 
    };
\addlegendentryexpanded{HC-B-60}
\end{axis}
\end{tikzpicture}
\caption{Time of breakthrough, as a function of the capillary number $Ca$, for different contact angles.}
\label{fig:t-durchbruch-vs-Ca-90vs60-HC-B}
\end{figure}
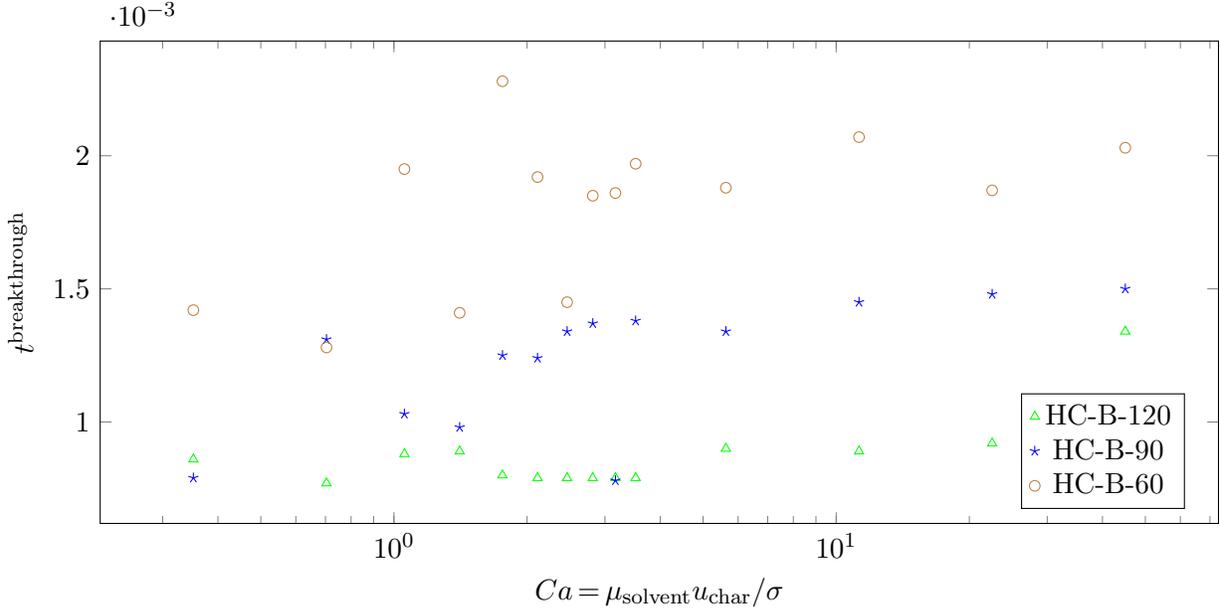

While in the previous section a neutral wettability and thus a contact angle of \ang{90} is assumed, the effect of a change in the contact angle is subsequently investigated. 
While no studies on the influence of surface wettability are available for the drying process, the effects of contact angle variations in the electrolyte filling process step in battery production are already known~\cite{shodiev2021insight}.
Since the model used in the present work can cover wetting effects, the impact of the contact angle during the drying process is accessible.
For this purpose, the results with a contact angle of \ang{90} are compared with contact angles of \ang{60} and \ang{120}, which indicate hydrophobic and hydrophilic behavior, respectively. 
\\
\\
Fig.~\ref{fig:deltah-vs-ca-contact} illustrates the effect of the contact angle on the normalized height difference $ \auswertungDeltah$ as a function of the capillary number $Ca$.
The different angles \ang{60}, \ang{90}, and \ang{120} are shown as symbols in the colors brown, blue, and green, respectively. 
For a contact angle of \ang{120}, no dependence of the height on the capillary number can be seen, whereas the values for a contact angle of \ang{60} and \ang{90} are scattered.
All values for a contact angle of \ang{60} are smaller than the values for the other contact angles. 
Without taking into account the capillary number $Ca=3.168$, this behavior is also evident between the contact angles \ang{90} and \ang{120}.
Here, the results are either identical or the values for a contact angle of \ang{90} are lower than those for \ang{120}. 
Compared to the results for an angle of \ang{90}, however, the results for an angle of \ang{60} are more widely distributed and depend on the capillary number.  
This parameter study indicates that as the contact angle decreases, the pinning on the surface is less pronounced, following that the normalized height difference is smaller. 
Such a behavior leads to a more uniform drying front with a decreasing contact angle, while at the same time the values are more scattered. 
While a constant value is obtained at a contact angle of \ang{120}, a large scatter is observed at a contact angle of \ang{60}.    
Fig.~\ref{fig:t-durchbruch-vs-Ca-90vs60-HC-B} shows the time of breakthrough for the three different angles, which are represented in brown, blue, and green symbols for the angles \ang{60}, \ang{90}, and \ang{120}, as a function of the capillary number $Ca$. 
The behavior of the breakthrough times is different for all three contact angles. 
In addition, the distribution of the data points assigned to the three different contact angles differs in the scattering property, with the variation being less pronounced at larger contact angles. 
At a contact angle of \ang{120}, the breakthrough time is almost constant. 
Wrapping up, the simulation studies point out that the contact angle plays an important, non-negligible role for pore emptying during the drying of battery electrodes. 
Furthermore, a tailored regulation of the contact angle offers potential for optimizing the drying process.  
This can be achieved by a special treatment of the particle surface, for example, by altering the surface structure. 
\subsection{Discussion of the simplifying assumption}
Both the model and the problem setup for the film drying simulations are based on assumptions and simplifications that are critically discussed in the following section, while at the same time, possible paths for future developments are identified.

\paragraph{Navier-Stokes Equations} A constant viscosity is chosen for the fluid, which represents the solvent and the additives. 
However, in general, the viscosity depends on the temperature and the shear rate.
In particular, the dependence of the viscosity on the shear rate is well known in the drying of battery electrodes~\cite{klemens2023process}. 
This dependence can be introduced by defining the shear rate $\dot{\gamma} = \sqrt{4 I_2(\boldsymbol{D})}$ based on the second basic invariant $I_2=\trace{\boldsymbol{D}\boldsymbol{D}}$ of the symmetric velocity gradient ~$\boldsymbol{D}$. 
For such an approach, it is necessary to model the viscosity as a function of the shear rate, $\mu=\mu(\dot{\gamma})$, which can be achieved by using a power law, for example (cf.~\cite{Ruzicka2013}).

\paragraph{Interpolation} In the simulations, the density of the air phase is 4 times higher than the real air density.
This is due to the fact that the interpolation of the physical parameters in the diffuse interface can lead to problems with high contrasts between the phases~\cite{Abels2012}.
The scaling can be justified by dimensional analysis.
The very small Reynolds and Weber number regarding the air phase indicate that only minimal inertia effects are present, which is also the case after scaling.
Therefore, the effect of the air velocity on the movement in the solvent is negligible, while the movement in the solvent is the relevant feature to be covered by the simulations.
Similarly, the air stream utilized for drying is not included in the problem setup, which is a widely used simplification~\cite{wolf2024computational}.
To solve the density interpolation problem, a model extension can be made by adding another term to the Navier-Stokes equation that accounts for the momentum exchange due to molecular diffusion, as proposed by Abels et al.~\cite{Abels2012}.

\paragraph{Evaporation}In the proposed model, the evaporation rate is modeled as a constant value.
Furthermore, the evaporation rate is evenly distributed over the diffuse interface. 
In reality, the saturation of the gas phase in the pores reduces the driving force for diffusive solvent transport in the gas phase, and the local evaporation rate can change.
One way to describe the drying rate $\Dot{m}_{\text{S}}$ is according to \cite{kumberg2019drying}: 
\begin{equation}
    \Dot{m}_{\text{S}}= \Tilde{M}_{\text{S}}\Tilde{\rho}_{\text{G}}\gamma_{\text{S,G}}\text{ln}\frac{(1-\Tilde{\gamma}_{\text{S,$\infty$}})}{(1-\gamma_{\text{S,Ph}})},
\end{equation}
where $\Tilde{M}_{\text{S}}$ is the molar mass of the solvent, $\Tilde{\rho}_{\text{G}}$ represents the molar density of the surrounding gas stream, $\beta_{\text{S,G}}$ is the mass transfer coefficient, $\Tilde{\gamma}_{\text{S,$\infty$}}$ is the molar fraction of the solvent in the drying air, and $\gamma_{\text{S,Ph}}$ denotes the molar fraction of the solvent in the solvent-gas interface.
This requires an additional transport equation that takes into account the solvent concentration in the gas phase.
As some of the parameters, such as the molar fractions, depend on the temperature, the temperature must also be taken into account.

The simplifications in the current work can be addressed by incorporating extensions to the discussed model, thereby enhancing the accuracy of the results.
Despite these simplifications, the presented model covers the most relevant physical effects for pore emptying during drying and requires fewer assumptions, compared to other models used to simulate film drying.

\section{Summary and outlook}
\label{sec:summary_outlook}
In this work, a full-field approach to simulate pore emptying in the second stage of battery electrode drying is presented for the first time.
This approach builds on the model proposed by Reder et al.~\cite{reder2022phase} and applies an Allen-Cahn approach as an alternative to the Cahn-Hilliard approach. 
Additionally, this approach takes into account the phase transition due to evaporation.
The model is validated using three different examples, which concern the macroscopic contact angle behavior during evaporation, the rise height of a fluid in a capillary, and the behavior of an evaporating fluid in a pore network. 
With respect to all validation cases, the model shows good agreement with the theoretical predictions.
Subsequently, the influence of the contact angle, viscosity, and microstructure on the drying behavior is investigated using realistic hard carbon microstructures as initial conditions.
For both microstructures, a correlation between viscosity and breakthrough time can be derived, which consists of the fact that the time to breakthrough also increases with increasing viscosity. 
However, this correlation can be negated as a result of the existence of privileged capillaries, by changing the contact angle. 
Two conclusions can be drawn from the studies, firstly that models which model the drying of battery electrodes should take the studied parameters into account, and secondly the production of battery electrodes can be optimized by changing the contact angle due to surface optimization.
A subsequent study will be conducted in order to investigate the influence on more complex microstructures like multilayer microstructures or 3D microstructures.
In particular, the simulation of 3D microstructures is important to determine the statistical effects of 2D structures and to correctly map the complexity of effective capillaries in all three spatial directions.
Additionally, the model can be easily extended to evaluate the binder behavior under drying conditions, since the full-field information of the solvent velocity is already known. 
For this purpose, a diffusion-convection equation needs to be integrated into the model. 
We hope that these extensions will allow us to better understand and optimize the drying of battery electrodes.
\begin{acknowledgments}
This work contributes to the research performed at CELEST (Center for
Electrochemical Energy Storage Ulm-Karlsruhe) and was funded by the
German Research Foundation (DFG) under Project ID 390874152 (POLiS
Cluster of Excellence). Within the cooperation of the cluster we thank Dr. Marcus Müller (Institute for Applied Materials IAM-ESS) for support regarding the SEM measurements. Support regarding the model development is provided through funding by the Helmholtz association within the programme "MTET", no. 38.02.01, which is gratefully acknowledged.
%
\\
\\
The authors declare no conflict of interest
\end{acknowledgments}

\appendix

\section{Calculation of the height of a capillary}
\label{sec:analytisch-steig}
\begin{figure}
\centering
    \includegraphics[scale =0.35]{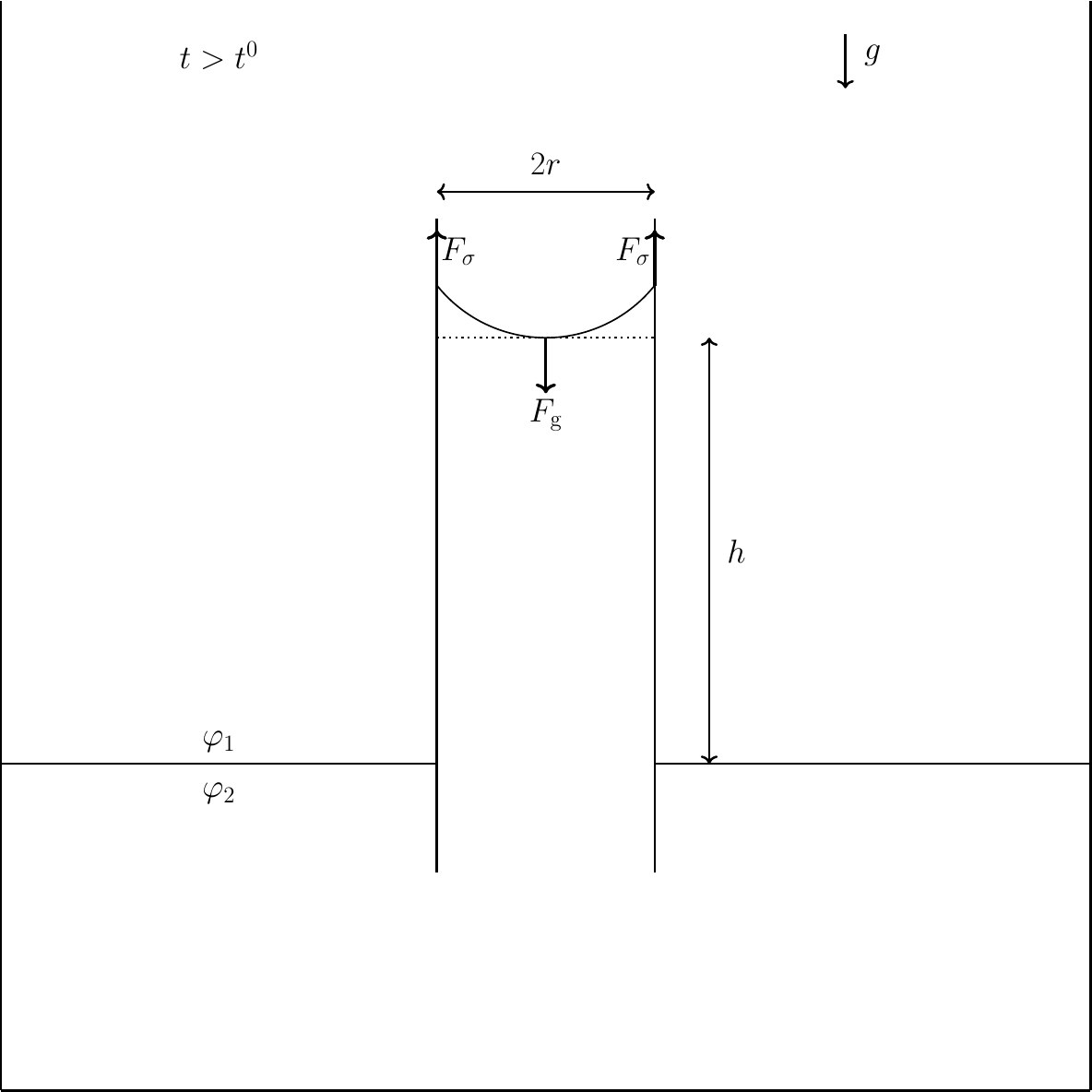}
     \caption{Balance of forces for a capillary.}
     \label{fig:GGW}
\end{figure}

The forces acting in a capillary are shown in Fig. \ref{fig:GGW}. 
The height of rise $h$ in the capillary in steady state can be calculated using a force balance. Thus, the gravitational force $F_{\text{g}}$ and the force $F_{\sigma}$ due to the surface tension on the wall balance each other out. The gravitational force is described by the relationship
\begin{align}
    F_{\text{g}} = (\rho_{2}-\rho_{1})  g  V,
\end{align}
where $\rho_{2}$ is the density of fluid 2 and $\rho_{1}$ is the density of fluid 1. $V$ denotes the volume of fluid 2 and $g$ describes the acceleration due to gravity. The surface tension force is given by
\begin{align}
 F_{\sigma} = \sigma \cos{\Theta} D,    
\end{align}
where $d$ represents the depth and $\Theta$ is the contact angle. The volume $V = A D$ of the liquid $\varphi_{2}$ is given  with the surface $A$ of fluid 2, which is parameterized via 
\begin{align}
    A &= 2 r D \left(h + \frac{r}{2 \cos(\Theta)} \left(2 -\sin(\Theta) - \frac{\arcsin(\cos(\Theta))}{\cos(\Theta)} \right) \right)
\end{align}
 and the depth $D$. Using the force balance $F_{\text{g}} = F_{\sigma}$, we obtain the rise height
\begin{align}
    h &= \frac{\sigma \cos{\Theta}}{(\rho_{2}-\rho_{1}) g r} -\frac{r}{2 \cos(\Theta)} \left(2 -\sin(\Theta) - \frac{\arcsin(\cos(\Theta))}{\cos(\Theta)} \right).
\end{align}
If the density contrast is high, the lower density can be neglected. This leads to the correlation
\begin{align}
    h &= \frac{\sigma \cos{\Theta}}{\rho_{2} g r} -\frac{r}{2 \cos(\Theta)} \left(2 -\sin(\Theta) - \frac{\arcsin(\cos(\Theta))}{\cos(\Theta)} \right),   
\end{align}
which is typically found in the literature~\cite{grunding2020comparative}.

\section{Volume conservation of the Allen-Cahn equation}
\label{volume-perserving-allen-cahn}
\begin{figure}[htbp]
\centering
\begin{subfigure}[c][][t]{0.3\textwidth}
\centering
\hspace{-1.65cm}
\includegraphics[scale =1]{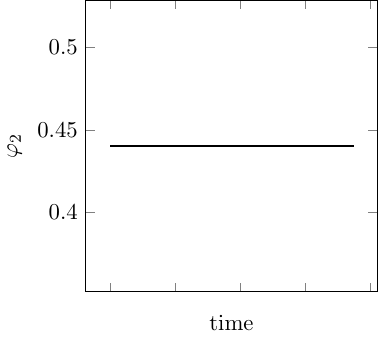}
\end{subfigure}
\begin{subfigure}[c][][t]{0.3\textwidth}
\centering
\vspace{-0.8cm}
			\begin{tikzpicture}[baseline]
			\begin{axis}[xlabel=, ylabel=,
				 xmin=0, xmax=1, xticklabels={},
				 ymin=0, ymax=1, yticklabels={},
				scale only axis=true,
                xlabel={},
				ticks=none,
				width=1\textwidth, height=1.0\textwidth,
				]
                \addplot graphics[xmin = 0.0, xmax = 1.0, ymin = 0, ymax = 1]{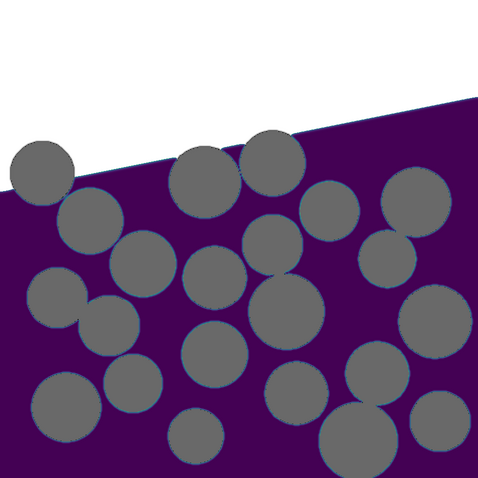};
		\end{axis}
\end{tikzpicture}

\end{subfigure}
\begin{subfigure}[c][][t]{0.3\textwidth}
\vspace{-0.8cm}
			\begin{tikzpicture}[baseline]
			\begin{axis}[xlabel=, ylabel=,
				 xmin=0, xmax=1, xticklabels={},
				 ymin=0, ymax=1, yticklabels={},
				scale only axis=true,
				ticks=none,
				width=1\textwidth, height=1.0\textwidth,
				]
                \addplot graphics[xmin = 0.0, xmax = 1.0, ymin = 0, ymax = 1]{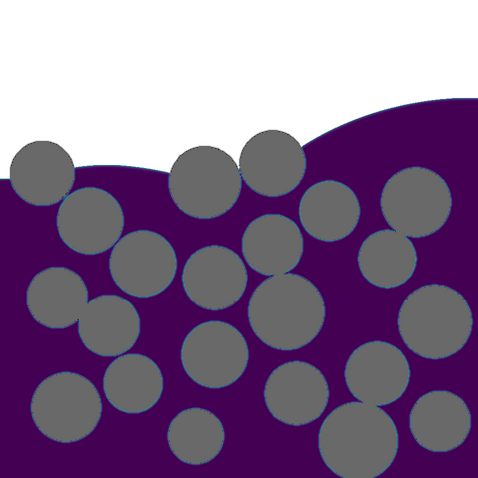};
		\end{axis}
\end{tikzpicture}
\end{subfigure}
\caption{Volume fraction of fluid $\varphi_2$ as a function of time (left), initial setting (middle) and final microstructure (right).}
\label{fig:phi-norm-final-refi-3}
\end{figure}

In this work, an Allen-Cahn approach is employed for the phase evolution. An Allen-Cahn approach without modifications is not volume-preserving. If no additional evaporation velocity is considered, the surface terms lead to a volume change due to curvature minimization. However, the present approach leads to a volume-preserving Allen-Cahn equation, if the factor $\kappa$ is set to zero. In this case, the term in equation~\eqref{eq:volume-velocity} gives an interfacial velocity that counteracts the volume shrinkage due to the curvature minimization dynamics. For $\kappa=0$, the present model therefore corresponds to volume-preserving approaches, as in Nestler et al.~\cite{nestler2005multicomponent}, using a Lagrange multiplier. Such an approach enforces the condition $\frac{d}{\text{d}t}\int \Tilde{\varphi}\text{d}t=0$. It is to be noted, that the curvature correction term~\cite{Sun2007} removes the curvature-minimizing dynamics and thus the corresponding volume change. 
However, in a numerical implementation this is only achieved approximately. Thus, a Lagrange multiplier is still required to ensure exact volume preservation. In this section, the expected behavior is to be validated. The initial values of the simulation can be found in Sec.~\ref{sec:validation}.
Initially, three phases are set, with the solids colored gray and the fluids white and blue. 
In Fig.~\ref{fig:phi-norm-final-refi-3} on the left, the volume fraction of $\varphi_2$ (blue) is shown as a function of time for $480\times480$ cells.
The middle and right images correspond to the initial and final states. The volume fraction is constant, while the surface changes according to the curvature minimization.

\section{Contact angle behavior of a static case}
 \label{contact-angle-static}
\begin{figure}[htbp]
    \centering
         \includegraphics[scale =0.45]{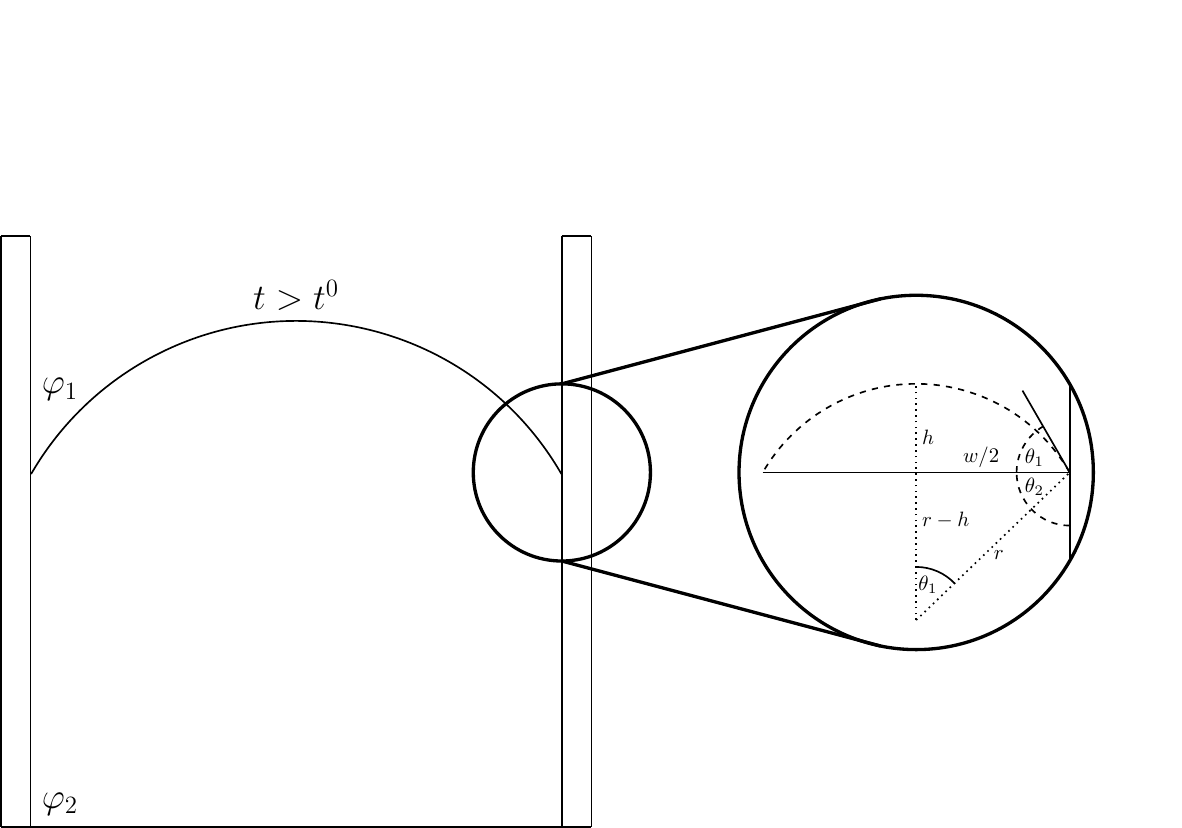}
    \caption{Schematic drawing of the fluid distribution at any time $t>t^0$, with geometric relationships at the three-phase interface between the two fluids and the capillary wall.}
         \label{fig:kontaktwinkel-tn}
\end{figure}

\begin{figure}[h!]
\centering
\includegraphics[scale =1]{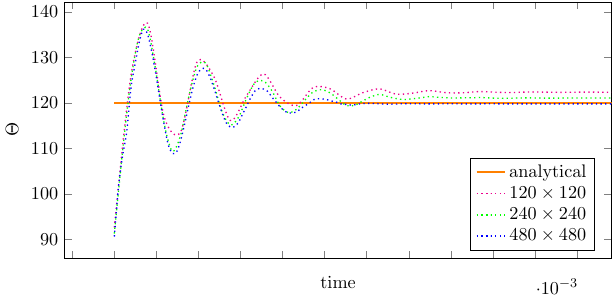}
\caption{Grid refinement for the static case with a contact angle of $60^{\circ}$.}
\label{fig:statisch-60}
\end{figure}

Fig.~\ref{fig:kontaktwinkel-tn} sketches the development of the fluids for a point in time $t>t^0$. Due to the balance of forces between the surface tension and the volume force due to gravity, an angle $\Theta$ is set at the three-phase interface in equilibrium. This angle can be calculated using the angular relationships
\begin{align}
    \frac{2h}{w} = \tan{(\frac{\theta_1}{2})}.
\end{align}
The parameters from Sec.~\ref{sec:dynamic-contanct-angle} are used for this simulation study, excluding $\kappa$ which is set to zero. Fig.~\ref{fig:statisch-60} shows the setting angle $\Theta=\theta_1+\theta_2$ as a function of time for different resolutions with $\theta_2 = \ang{90}$. The resolutions are $120\times120$, $240\times240$, and $480\times480$ for $n_{\text{x}}\times n_{\text{y}}$. 
All three resolutions follow the same behavior by starting with an initial angle of \ang{90} and proceeding with an oscillation around the respective stationary solution before approaching it. 
With increasing refinement of the resolution, the deviation from the theoretical contact angle decreases. The values in equilibrium are  $\Theta=122.36$, $\Theta=121.08$ and $\Theta=119.81$, from the coarsest to the finest resolution.  
This simulation study reveals a converging behavior of the model capable to reproduce the wetting behavior for a stationary interface. The wetting behavior of a moving interface, taking evaporation into account, is found in~\eqref{sec:dynamic-contanct-angle}.

\section{Derivation of the diffuse wetting boundary condition for the Allen-Cahn approach}
\label{derivation-diffuse-wetting-bc}
To impose the moving boundary condition at the solid-fluid interface, we follow the approach of Li et al.~\cite{li2009solving}. The considered phase-field equation~\eqref{eq:Allen-Cahn-Gleichung-Evap} is given by
\begin{align}
  \Dot{\Tilde{\varphi}} &=   M \left(\beta \partial_{\Tilde{\varphi}}\psi - \alpha  \left(\grad^2 \Tilde{\varphi} - ||\grad\Tilde{\varphi}||\grad \cdot \frac{\grad\Tilde{\varphi}}{||\grad\Tilde{\varphi}||}\right)\right) + v^{\text{e}}||\grad \Tilde{\varphi}||, \qquad \bm{x} \in V
  \label{eq:fieldeq-anhang}
\end{align}
with the boundary conditions~\eqref{eq:wettingBC} 
\begin{align}
    \alpha \grad \Tilde{\varphi} \cdot \bm{n} &= (\sigma_{\text{2s}}-\sigma_{\text{1s}})\partial_{\Tilde{\varphi}}h^{\text{ff}}(\Tilde{\varphi}), \qquad \bm{x} \in \partial V.
    \label{eq:boundary-anhang}
\end{align}
First, we  multiply the equation by the test function $\Psi$.  
Then we integrate over the fluid volume $V \subset \Omega$, with $\Omega$ as the whole domain. 
This gives the weak form of the equation and is represented by 
\begin{align}
     0=\intVol{V}{ \Psi  \Dot{\Tilde{\varphi}} - M \Psi \beta \partial_{\Tilde{\varphi}}\psi  + M \Psi \alpha \left(\grad^2 \Tilde{\varphi} - ||\grad\Tilde{\varphi}||\grad \cdot \frac{\grad\Tilde{\varphi}}{||\grad\Tilde{\varphi}||}\right) - \Psi v^{\text{e}} ||\grad \Tilde{\varphi}|| }.
\end{align}
Taking into account the Gauss theorem, we obtain
\begin{align}
\begin{split}
    0&=\intVol{V}{ \Psi \Dot{\Tilde{\varphi}} - M \Psi \beta \partial_{\Tilde{\varphi}} \psi - M \alpha \left( \grad \Tilde{\varphi} \cdot \grad \Psi - ||\grad\Tilde{\varphi}||\grad \cdot \frac{\grad\Tilde{\varphi}}{||\grad\Tilde{\varphi}||}\right)  - \Psi v^{\text{e}} ||\grad \Tilde{\varphi}|| } \\ &\qquad +\intSurf{\partial V}{ \alpha M  \Psi \grad \Tilde{\varphi} \cdot \bm{n}}.
    \label{eq:after-gauss}
 \end{split}   
\end{align}
Introducing the indicator function $I$ and the Dirac distribution  $\delta_{\Gamma}$, the equation \ref{eq:after-gauss} can be written as an integral of the whole domain
\begin{align}
\begin{split}
    0&=\intVol{\Omega}{ I \Psi \Dot{\Tilde{\varphi}} - M\Psi \beta I \partial_{\Tilde{\varphi}} \psi -M \alpha I  \left( \grad \Tilde{\varphi} \cdot \grad \Psi - ||\grad\Tilde{\varphi}||\grad \cdot \frac{\grad\Tilde{\varphi}}{||\grad\Tilde{\varphi}||}\right) \\ &\qquad
    - I \Psi v^{\text{e}} ||\grad \Tilde{\varphi}|| + M\delta_{\Gamma}\alpha \Psi \grad \Tilde{\varphi} \cdot \bm{n} }.
    \end{split}
\end{align}
Using the context
\begin{align}
    \intVol{\Omega}{ \grad \cdot  (I \Psi \grad \Tilde{\varphi})} = \intVol{\Omega}{ \Psi \grad \cdot (I \grad \Tilde{\varphi})} + \intVol{\Omega}{I \grad \Tilde{\varphi} \cdot \grad \Psi}
\end{align}
we receive,
\begin{align}
\begin{split}
        0&=\intVol{\Omega}{ \Psi (I \Dot{\Tilde{\varphi}}  + M \alpha  \grad \cdot(I  \grad\Tilde{\varphi}) -M \alpha I \left( ||\grad\Tilde{\varphi}||\grad \cdot \frac{\grad\Tilde{\varphi}}{||\grad\Tilde{\varphi}||}\right) \\ &\qquad 
        - M \beta I \partial_{\Tilde{\varphi}} \psi - I v^{\text{e}} ||\grad \Tilde{\varphi}|| + M \alpha \delta_{\Gamma} \grad \Tilde{\varphi} \cdot \bm{n}) }.
        \end{split}
\end{align}
By applying the fundamental lemma of variation, we  get for $ \bm{x} \in \Omega$:
\begin{align}
\begin{split}
    I  \Dot{\Tilde{\varphi}} &=  M\left(-\alpha  \grad \cdot(I  \grad \Tilde{\varphi}) +M \alpha I \left( ||\grad\Tilde{\varphi}||\grad \cdot \frac{\grad\Tilde{\varphi}}{||\grad\Tilde{\varphi}||}\right) + \beta I \partial_{\Tilde{\varphi}} \psi - \alpha \delta_{\Gamma} \grad \Tilde{\varphi} \cdot \bm{n}\right) \\ &\qquad + I v^{\text{e}} ||\grad \Tilde{\varphi}||.
    \end{split}
\end{align}
Using the boundary condition in equation~\eqref{eq:wettingBC} to express the normal gradient $\grad\Tilde{\varphi} \cdot \bm{n}$, we obtain the following for $ \bm{x} \in \Omega$:
\begin{align}
   \begin{split}
        I  \Dot{\Tilde{\varphi}} &=  M\left(\beta I \partial_{\Tilde{\varphi}} \psi + M \alpha I \left( ||\grad\Tilde{\varphi}||\grad \cdot \frac{\grad\Tilde{\varphi}}{||\grad\Tilde{\varphi}||}\right) - \alpha  \grad \cdot(I  \grad \Tilde{\varphi})  - \delta_{\Gamma} (\sigma_{\text{2s}}-\sigma_{\text{1s}})\partial_{\Tilde{\varphi}}h^{\text{ff}}(\Tilde{\varphi})\right) \\ &\qquad + I v^{\text{e}} ||\grad \Tilde{\varphi}||.
        \end{split}
\end{align}
This sharp interface equation represents the boundary value problem, which consists of the field equation~\eqref{eq:fieldeq-anhang} and the boundary conditions~\eqref{eq:boundary-anhang}.
To transform the sharp interface model into the diffuse interface model, we approximate the indicator function by $I\approx h^{\text{fs}}(\varphi^{\text{f}})$ and the Dirac distribution as $\delta_{\Gamma} = || \grad I || \approx \partial_{\varphi^{\text{f}}}h^{\text{fs}}(\varphi^{\text{f}})||\grad\varphi^{\text{f}}|| $. Thus, we finally get 
\begin{align}
    \begin{split}
         h^{\text{fs}}  \Dot{\Tilde{\varphi}} &=  M\Big[\beta h^{\text{fs}} \partial_{\Tilde{\varphi}} \psi - \alpha \left(\grad \cdot (h^{\text{fs}} \grad \Tilde{\varphi}) - h^{\text{fs}} ||\grad\Tilde{\varphi}||\grad \cdot \frac{\grad\Tilde{\varphi}}{||\grad\Tilde{\varphi}||}\right)   \\ &\qquad
            - \partial_{\varphi^{\text{f}}}h^{\text{fs}}(\varphi^{\text{f}})||\grad\varphi^{\text{f}})|| (\sigma_{\text{2s}}-\sigma_{\text{1s}})\partial_{\Tilde{\varphi}}h^{\text{ff}}(\Tilde{\varphi})\Big] + h^{\text{fs}} v^{\text{e}} ||\grad \Tilde{\varphi}||, \qquad  \bm{x} \in \Omega. 
             \end{split}        
\end{align}

\section{Derivation of the theoretical volume fraction of the solvent phase}
 \label{derivation-of-evap}
According to Jaiser et al. \cite{jaiser2016investigation}, the drying rate during the constant rate period (CRP) can be calculated using the equation
\begin{align}
 \Dot{r}_{\text{solvent}} = \left.\frac{\text{d}X}{\text{d}t}\right|_{\text{CRP}}  M_{\text{S}}.
\end{align}
 Here, $\Dot{r}_{\text{solvent}}$ is the area-specific evaporation rate in \si{\kg\per\square\metre\second}, $X$ is the ratio of solvent to solid masses, with $X=\frac{m_{\text{solvent}}}{m_{\text{solid}}}$, $t$ is the time in \si{\second}, and $M_{\text{S}}$ is the area-specific weight of the dry film in \si{\kg\per\square\metre}. By integrating the equation and using the masses, we obtain 
\begin{align}
 \frac{m_{\text{solvent}}}{m_{\text{solid}}} = \frac{m_{\text{solvent}}^0}{m_{\text{solid}}^0} - \frac{\Dot{r}_{\text{solvent}}}{M_{\text{S}}}t.
 \label{eq:theo_volume_mass}
\end{align}
Here, the index $0$ indicates the initial quantities. Equation \ref{eq:theo_volume_mass} can be expressed in terms of volume, using $V=m/\rho$, which results in the following for the volume of the solvent:
\begin{align}
 V_{\text{solvent}} = V_{\text{solvent}}^0 - \frac{\rho_{\text{solid}} V_{\text{solid}}}{\rho_{\text{solvent}} M_{\text{S}} } \Dot{r}_{\text{solvent}} t.
 \label{eq:theo_volume_vol}
\end{align}
 Introducing the variable $\chi_{\alpha} = \frac{V_{\alpha}}{V^{0}_{\text{film}}}$ as the volume fraction of a phase $\alpha$, related to the film volume $V^{0}_{\text{film}}=V_{\text{solvent}}^0 + V_{\text{solid}}$, we obtain
\begin{align}
 \chi_{\text{solvent}} = - \frac{\rho_{\text{solid}} \chi_{\text{solid}}}{\rho_{\text{solvent}} M_{\text{S}} } \Dot{r}_{\text{solvent}} t + \chi_{\text{solvent}}^0,
\end{align}
 by dividing the equation \ref{eq:theo_volume_vol} by the volume. By introducing the abbreviation
\begin{align}
 \kappa = \frac{\rho_{\text{solid}} \chi_{\text{solid}}}{\rho_{\text{solvent}} M_{\text{S}} } \Dot{r}_{\text{solvent}},
 \label{eq:kappa-and-evaprate}
\end{align}
 we finally obtain the relationship
\begin{align}
 \chi_{\text{solvent}} = - \kappa t + \chi_{\text{solvent}}^0,
\end{align}
 with \si{\per\second} as unit for $\kappa$.
\section{Results for the $\hat{x}$-coordinate}
\label{coordinate} 
\begin{figure}[htbp]
\centering
\begin{subfigure}[c]{1\textwidth}
\centering
\includegraphics[scale =1]{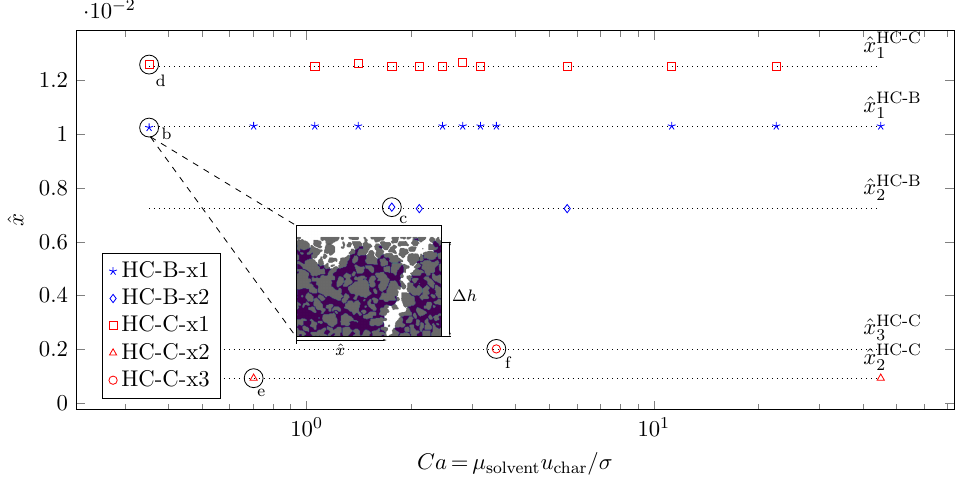}
\subcaption{Non-dimensionalized $\hat{x}$-coordinate of the breakthrough point, as a function of the capillary number $Ca$ for different microstructures. Data points relating to the HC-B and HC-B microstructure are represented by blue and red symbols. The horizontal lines represent the different clusters.}
\label{fig:x-durchbruch-vs-Ca}
\end{subfigure}
\begin{subfigure}[c]{0.19\textwidth}
\setlength{\fboxsep}{0cm}
\setlength{\fboxrule}{0.03cm}
\centering
\fbox{\includegraphics[scale=0.13]{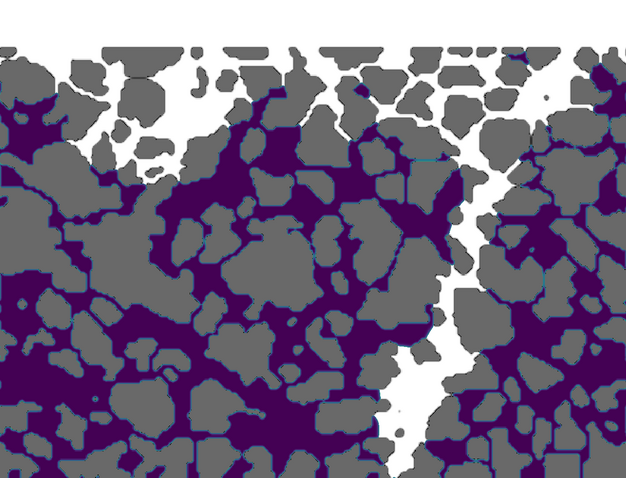}} 
\subcaption{HC-B-x1 $(\Ca=\num{0.35})$}
\label{fig:HC-B-x1}
\end{subfigure} 
\begin{subfigure}[c]{0.19\textwidth}
\setlength{\fboxsep}{0cm}
\setlength{\fboxrule}{0.03cm}
\centering
\fbox{\includegraphics[scale=0.13]{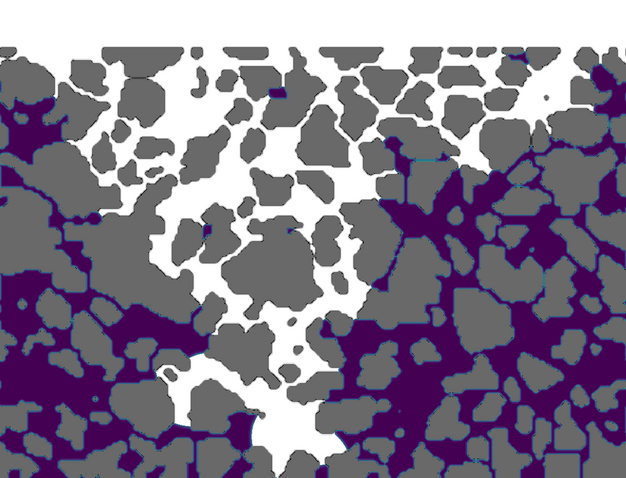}}  
\subcaption{HC-B-x2 $(\Ca=\num{1.76})$}
\label{fig:HC-B-x2}
\end{subfigure} 
\begin{subfigure}[c]{0.19\textwidth}
\setlength{\fboxsep}{0cm}
\setlength{\fboxrule}{0.03cm}
\centering
\fbox{\includegraphics[scale=0.13]{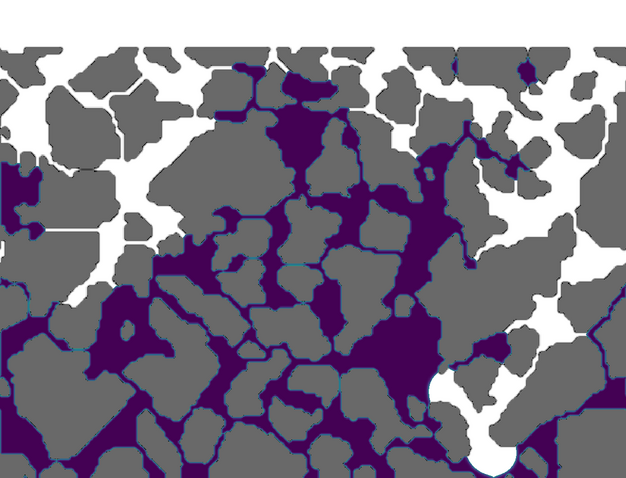}}  
\subcaption{HC-C-x1 $(\Ca=\num{0.35})$}
\label{fig:HC-C-x1}
\end{subfigure} 
\begin{subfigure}[c]{0.19\textwidth}
\setlength{\fboxsep}{0cm}
\setlength{\fboxrule}{0.03cm}
\centering
\fbox{\includegraphics[scale=0.13]{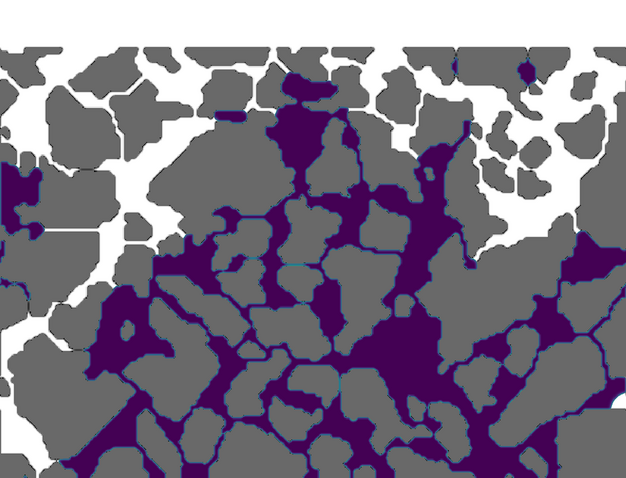}}  
\subcaption{HC-C-x2 $(\Ca=\num{0.70})$}
\label{fig:HC-C-x2}
\end{subfigure} 
\begin{subfigure}[c]{0.19\textwidth}
\setlength{\fboxsep}{0cm}
\setlength{\fboxrule}{0.03cm}
\centering
\fbox{\includegraphics[scale=0.13]{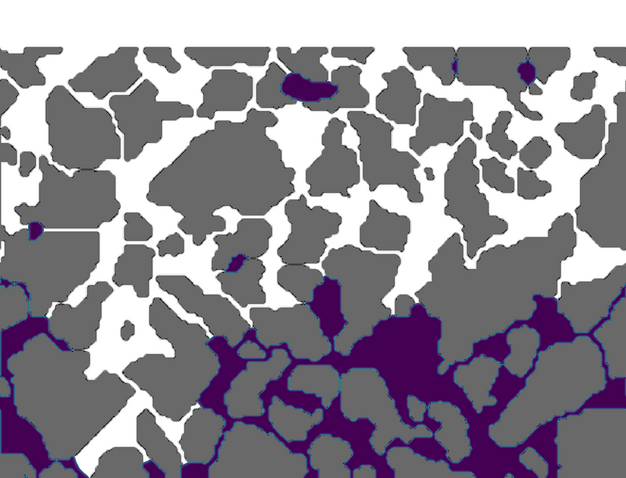}}  
\subcaption{HC-C-x3 $(\Ca=\num{3.52})$}
\label{fig:HC-C-x3}
\end{subfigure}
\caption{Dependence of the non-dimensionalized $\hat{x}$-coordinate of the breakthrough point on the capillary number $Ca$ (top). Simulation results for selected combinations of capillary numbers and $\hat{x}$-coordinates of the breakthrough point (bottom). The solvent is blue, the air is white, and particles are shown in gray.}
\end{figure}

Fig.~\ref{fig:x-durchbruch-vs-Ca} depicts the non-dimensionalized $\hat{x}$-coordinate of the breakthrough point as a function of the capillary number for the microstructures HC-B and HC-C in blue and red, respectively. The $\hat{x}$-coordinates assume identical values for different capillary numbers. 
Figs.~\ref{fig:HC-B-x1}-\ref{fig:HC-C-x3} illustrate the simulated microstructures at the time of breakthrough. Fig. \ref{fig:HC-B-x1} and \ref{fig:HC-B-x2} refer to the HC-B microstructure, for the capillary numbers $\Ca=\num{0.35}$ and $\Ca=\num{1.76}$. A different $\hat{x}$-value can be observed for different capillary numbers. This is characterized by a differently favored capillary that forms until the breakthrough. The microstructure HC-C is given in the Figs.~\ref{fig:HC-C-x1} to~\ref{fig:HC-C-x3} for the capillary numbers $\Ca=\num{0.35}$, $\Ca=\num{0.70}$, and $\Ca=\num{3.52}$. All three capillary numbers supply different values for $\hat{x}$ and thus differently favored capillary pathways. Figs. \ref{fig:HC-C-x1} to \ref{fig:HC-C-x2} differ slightly with respect to the height difference $\Delta h$. In summary, a significant influence of the microstructure and the viscosity, represented by a different capillary number, on the drying behavior can be observed.

\bibliography{literatur_mw}

\end{document}